\documentclass[%
 aip,
 amsmath,amssymb,
 reprint,%
]{revtex4-1}
\usepackage{hyperref}
\usepackage{graphicx}% Include figure files
\usepackage{dcolumn}% Align table columns on decimal point
\usepackage{bm}% bold math
\usepackage[utf8]{inputenc}
\usepackage[T1]{fontenc}
\usepackage{mathptmx}
\usepackage{etoolbox}
\usepackage{soul}
\usepackage{xcolor}
\makeatletter
\def\@email#1#2{%
 \endgroup
 \patchcmd{\titleblock@produce}
  {\frontmatter@RRAPformat}
  {\frontmatter@RRAPformat{\produce@RRAP{*#1\href{mailto:#2}{#2}}}\frontmatter@RRAPformat}
  {}{}
}%
\makeatother
\begin{document}

\preprint{AIP/123-QED}

\title{Microscopic Theory of Nonlinear Rheology and Double Yielding in Dense Attractive
 Glass Forming Colloidal Suspensions}
% Force line breaks with \\
\author{Anoop Mutneja}
\affiliation{Departments of Materials Science and Engineering, University of Illinois, Urbana, Illinois, 61801, USA}
\affiliation{Materials Research Laboratory, University of Illinois, Urbana, Illinois, 61801, USA}
\author{Kenneth S.Schweizer}%
 \email{kschweiz@illinois.edu.}
\affiliation{Departments of Materials Science and Engineering, University of Illinois, Urbana, Illinois, 61801, USA}
\affiliation{Materials Research Laboratory, University of Illinois, Urbana, Illinois, 61801, USA}
\affiliation{Departments of Chemistry, University of Illinois, Urbana, Illinois, 61801, USA}
\affiliation{Departments of Chemical and Biomolecular Engineering, University of Illinois, Urbana, Illinois, 61801, USA}

\date{\today}% It is always \today, today,

\begin{abstract}
Yielding of amorphous glasses and gels is a mechanically driven transformation of a material from the solid to liquid state on the experimental timescale. It is a ubiquitous fundamental problem of nonequilibrium physics of high importance in material science, biology, and engineering applications such as processing, ink printing, and manufacturing. However, the underlying microscopic mechanisms and degree of universality of the yielding problem remain theoretically poorly understood. We address this problem for dense Brownian suspensions of nanoparticles or colloids that interact via repulsions that induce steric caging and tunable short range attractions that drive physical bond formation. In the absence of deformation, these competing forces can result in fluids, repulsive glasses, attractive glasses, and dense gels of widely varying elastic rigidity and viscosity. Building on a quiescent microscopic theoretical approach that explicitly treats attractive bonding and thermally-induced activated hopping, we formulate a self-consistent theory for the coupled evolution of the transient and steady state mechanical response, and structure as a function of stress, strain, and deformation rate over a wide range of high packing fractions and attraction strengths and ranges. Depending on the latter variables, under step rate shear the theory predicts three qualitatively different transient responses: plastic-like (of two distinct types), static yielding via a single elastic-viscous stress overshoot, and double or 2-step yielding due to an intricate competition between deformation-induced bond breaking and de-caging. A predictive understanding of multiple puzzling experimental observations is achieved, and the approach can be extended to other nonlinear rheological protocols and soft matter systems.
\end{abstract}

\maketitle
\section{Introduction}
The microscopic mechanism of yielding in ultra-dense suspensions of nanoparticles or colloids (and their mixtures) is a problem of high scientific importance and of wide materials and engineering importance \cite{BonnRevModPhys2017,JoshiRheoActa2018,Berthier2024,Wagner_Mewis_2021,Joshi2020}. Its fundamental understanding at a predictive microscopic level remains a major challenge for nonequilibrium statistical mechanics. Yielding can be viewed as a mechanically-driven solid-to-liquid transition that can occur under diverse rheological protocols and via qualitatively different mechanisms due to the wide tunability of particle softness and shape, interparticle forces, and particle concentrations that determine emergent solidity \cite{BonnRevModPhys2017,JoshiRheoActa2018,Berthier2024,Wagner_Mewis_2021,Joshi2020}. Amorphous solids can exist on the experimental timescale due to strong repulsive interactions that sterically localize particles (caging), which compete with short range attractions of diverse origin (e.g., grafted polymer brushes in poor solvents, van der Waals attractions, polymer-mediated entropic depletion attraction) that can induce long-lived transient physical bonding, resulting in richly varied linear and nonlinear mechanical properties \cite{Pham2008,Koumakis2011,Moghimi2020,Sciortino2005,AmannJOR2014,Wagner_Mewis_2021,Larson1998,Joshi2020}. How different dynamical constraints soften via external forces and mechanically-driven structural changes, and are eventually overcome in Brownian systems via deformation-assisted thermally activated processes that are unmeasurably long in equilibrium, is a major challenge to understand and is our focus here.\\

\begin{figure}
\includegraphics[width=0.30\textwidth]{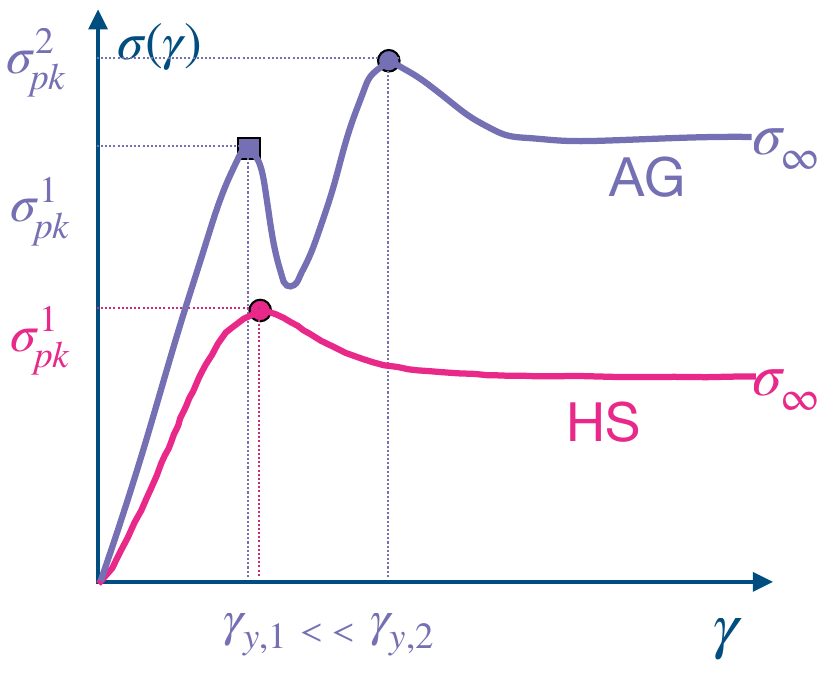}
	\caption{Schematic of single yielding versus double yielding in the representation of a stress ($\sigma$) vs strain ($\gamma$) curve at a fixed strain rate ($\dot{\gamma}$) which exhibits either a single overshoot (e.g., dense hard sphere suspensions) or a double overshoot (e.g., dense attractive glass (AG)). The overshoots are characterized by static yield stresses $\sigma_{pk}^1$ and $\sigma_{pk}^2$ at yield strains $\gamma_{y,1}$ and $\gamma_{y,2}$, respectively. The extremely ductile nature of the second yield process (very large yield strain) observed in experiments on attractive glasses \cite{Pham2008,Koumakis2011,Moghimi2020} is indicated. 
    } 
 \label{fig1}
\end{figure}

A classic transient rheological experiment to probe such soft matter systems is step rate shear (sometimes called “continuous startup shear”) at a fixed deformation rate ($\dot{\gamma}$) initiated at zero time ($t=0$) where stress ($\sigma$) is measured as a function of accumulated strain defined as $\gamma\equiv\dot{\gamma}t$. Typically, four distinct regimes are observed \cite{Larson1998,KoumakisJOR2016,Wagner_Mewis_2021}: elastic, anelastic, overshoot, and steady-state, per the schematic Fig.\ref{fig1}. For dense particle suspensions (and granular materials \cite{FarainPRL2023}), shear stress initially grows linearly with strain per an elastic solid, followed by an anelastic sub-linear growth of a nonlinear elastic origin and/or via the onset of local dissipative activated relaxation processes triggered by deformation. At very long times or high accumulated strains, a steady state is achieved with stress saturating at a plateau $\sigma_\infty$, indicative of a flowing nonequilibrium fluid. \\

The intermediate strain regime defines the fascinating transient response of high practical relevance where the material transitions from solid-like to viscous-like. At least 3 qualitatively different behaviors can occur:  (i) ideal plastic flow where stress monotonically approaches from below the nonequilibrium state; (ii) a non-monotonic response characterized by a single overshoot (static yield point) indicating a crossover from solid-like to liquid-like behavior at a system-specific characteristic stress, strain, and overshoot amplitude (Fig.\ref{fig1}); (iii) a “double yielding” or “2 step” response for ultra-dense attractive particle suspensions that are both dynamically caged and form strong tight physical bonds characterized by two stress overshoots with nonuniversal features that depend on the microscopic interactions, packing fraction, and deformation rate (Fig.\ref{fig1}). For case (iii), qualitatively, the first overshoot at lower strain has been suggested to arise from deformation-induced bond weakening and breakage, while the second overshoot signifies cage rearrangement which can be very strongly modified per the surprising observation of extremely large (ductile) yield strains of $100\%$ \cite{Pham2008,Koumakis2011,Moghimi2020}.%\st{more akin to polymeric matter} \cite{doi1988theory,Larson1998,Roth2016}.
Despite interesting attempts based on ideal mode coupling theory (MCT) \cite{AmannJOR2014,Fuchs2002,FuchsFarday2002,Fuchs2005} largely focused on hard sphere and repulsive colloid glass forming suspensions, and an abstract entropy crisis landscape perspective proposed for double yielding in attractive glasses \cite{ZamponiPRL2018}, it appears that no successful microscopic theory exists for this two-step yielding process that explains its physical origins, nor how it can be driven in a multi-faceted manner by densification, increasing shear rate, changing attraction strength, and varying the nanometer length scale of physical bonding. The goal of the present article is to formulate such a theory and apply it to understand diverse experimental behaviors characteristic of dense attractive Brownian colloidal suspensions. 

Many different physical systems exhibit double yielding \cite{Wagner_Mewis_2021,Joshi2020}, including binary colloidal mixtures with large size disparity \cite{Sentjabrskaja2013}, magnetorheological fluids \cite{Magneto2014}, gels \cite{Moghimi2017,Ahuja2020,Shao2013}, clay pastes \cite{Shukla2015}, { emulsions \cite{Datta2011}} and capillary suspensions \cite{Ahuja2017} where the inter-cluster and intra-cluster physical bonds occur on two very different length scales. {Some are Brownian, some are non-Brownian or granular.} But the simplest and widely studied system that displays the full richness of double yielding is ultra-dense suspensions of Brownian sticky particles, often called “attractive glasses” (AG). Under quiescent conditions, they exhibit striking re-entrant phenomena such as weak attraction driven glass melting (Fig.\ref{fig3}(a)) \cite{Bergenholtz1999, Dawson2000, Zaccarelli2002, Pham2002, WGotze_2003, Kaufman2006, Zaccarelli2009, Willenbacher2011,Royall2018,Fullerton2020,Luo2021} that is associated with non-monotonic variation with attraction strength of single particle and collective structural relaxation times and the linear elastic shear modulus (Fig.\ref{fig3}(b)) \cite{Pham2008, Atmuri2012,Willenbacher2011}, and intermediate time sub-diffusive transport regimes \cite{Zaccarelli2008,Fullerton2020}. In this article, we focus on understanding such systems, though the proposed statistical mechanical ideas are general.\\

Qualitatively, we view the static yield overshoot phenomenon as a signature of the subtle competition between deformation-induced dynamic constraint softening of steric cages and physical bonds, and an activated solidity rebuilding process. As known for other amorphous materials such as polymer glasses \cite{Chen2011,Roth2016} that exhibit a \textit{single} yielding transition, such softening can occur via \textit{both} deformation-induced structural changes (sometimes referred to as “rejuvenation”) and external shear forces transduced to the particle scale where cages and physical bonds exist. The solidity rebuilding process associated with internal stress relaxation driven by deformation-dependent (typically activated) dynamics competes with this softening and attempts to drive the system back to equilibrium (ala an “aging” process). For the simplest hard sphere (HS) colloidal suspensions that form repulsive glasses, the overshoot magnitude exhibits a remarkable non-monotonic variation with strain rate and tends to vanish at ultra high packing fractions \cite{KoumakisPRL2012,KoumakisJOR2016}.\\

The striking rheological features mentioned above have only very recently been understood based on a microscopic nonequilibrium statistical mechanical theory \cite{GhoshJOR2023} that captures the above physics in a predictive manner by relating macroscopic stress to microscopic forces and the activated motion of individual particles. Significant support for the adopted simplifying microrheological perspective has been accumulating based on combined rheology, confocal microscopy, and diffusion measurements \cite{Larson1998,KoumakisJOR2016,LauratiJOP2012,AmannJCP2015,Marenne2017,Besseling2017,Denisov2013,Edera2024,Aime2023,Cipelletti2020}. Crucially, the coupled deformation modified activated relaxation process, stress relaxation, and structural deformation and relaxation are treated in a unified manner in ref.\cite{GhoshJOR2023}. If one ignores the nonequilibrium structural evolution, then steady state phenomena such as shear thinning can still be well captured, but the transient overshoot behavior is entirely missed \cite{GhoshJCP2020}.\\

There has been work by others to model phenomenologically the competing mechanisms described above with highly coarse grained models, with and without (granular, quasi-static deformation) thermal fluctuations, with an emphasis on an overshoot as a signature of macroscopic shear banding \cite{FieldingPRL2020,FieldingPRR2022} . This type of approach is not relevant to our work, not only because we address the problem microscopically, but because we analyze stress overshoots under homogeneous deformation conditions, and not as a casual consequence of an inhomogeneous constitutive instability. Indeed, the motivating experiments on colloid depletion attraction systems that we address \cite{Pham2008,Koumakis2011,Moghimi2020} do not report macroscopic shear banding, nor do they invoke aging, pre-shear history, or other effects that we do not consider here. For athermal granular systems, a phenomenological non-microscopic model has been recently developed \cite{FarainPRL2023} that includes in an elementary manner competing “aging” and “rejuvenation” processes under deformation that can qualitatively capture the experimentally observed strain-rate independent single overshoot behavior \cite{JoshiRheoActa2018}. \\

Here, we employ, conceptually synthesize, and qualitatively extend a suite of recently developed force-level statistical mechanical theories for the quiescent dynamics of dense attractive particle suspensions and the nonlinear rheology of ultra-dense Brownian HS suspensions to formulate what, to our knowledge, is the first successful microscopic theory for the nonlinear rheology of sticky particle fluids that can undergo simple plastic flow, single yielding, and double yielding. The developed approach requires the formulation of coupled mechanical constitutive and nonequilibrium structural evolution equations. The different consequences of repulsive and attractive forces (coupled caging and bonding) are \textit{explicitly} treated, which has been shown recently \cite{Dell2015,Ghosh2019,GhoshPRE2020,Mutneja2024} to be essential for properly capturing the rich quiescent re-entrant behaviors behavior observed in experiments \cite{Pham2002,Kaufman2006,Wagner_Mewis_2021} and simulations \cite{Zaccarelli2008,Fullerton2020}. The theory can address other rheological protocols such as stress-controlled creep, discontinuous step strain, and stress relaxation after an interrupted startup continuous deformation. This broad capability is illustrated by applying the present advance to the problem of “elastic yielding” in the absence of activated relaxation. The latter is germane to yielding immediately after the imposition of a discontinuous step-strain, and potentially to granular systems where thermally-induced hopping is not possible. Indeed, we show that our predictions for this limiting case are deeply related to the mechanism of double yielding of Brownian suspensions under {step rate} shear. \\ 

To place our new work in context, in Section \ref{Sec2} we present background information on the theoretical methods employed. Section \ref{Sec3} presents the general aspects of the new theory ideas for dense suspensions, including deformation-induced structure changes, activated relaxation, and elementary rheological consequences. Section \ref{Sec4} presents a comprehensive application to {step rate} rheology covering hard spheres, attractive glasses, and dense gels. The evolution of the rheological response with attraction strength and range, shear rate, and packing fraction are explored, and qualitative comparisons with experiments are briefly discussed. Connections with experiment and testable predictions are presented in Section \ref{Sec5}. Section \ref{Sec6} provides a conceptual overview of the ideas of the new theory that underlie the results presented in Sections \ref{Sec3} and \ref{Sec4}. The article concludes in Section \ref{Sec7} with a discussion and future outlook. The Supporting Information (SI) contains additional results that buttress the conclusions drawn in the main article. \\

\section{Theoretical Background and Methods}\label{Sec2}
To set the stage for our new work and for the benefit of the reader, we first review the background theoretical elements, the details of which are well documented in the literature \cite{Schweizer2005,Mirigian2014,Dell2015,GhoshJOR2023,Mutneja2024}. The key aspects are: (i) the microscopic Elastically Collective Nonlinear Langevin Equation (ECNLE) theory and dynamic free energy concept, the activated structural (alpha) relaxation time, the elastic modulus, and an explicit treatment of repulsive and attractive forces for quiescent liquids, and (ii) a generalized Maxwell model constitutive equation approach that includes the evolution of all relevant quantities with deformation.\\

\subsection{Quiescent fluids} 
\begin{figure}
\includegraphics[width=0.40\textwidth]{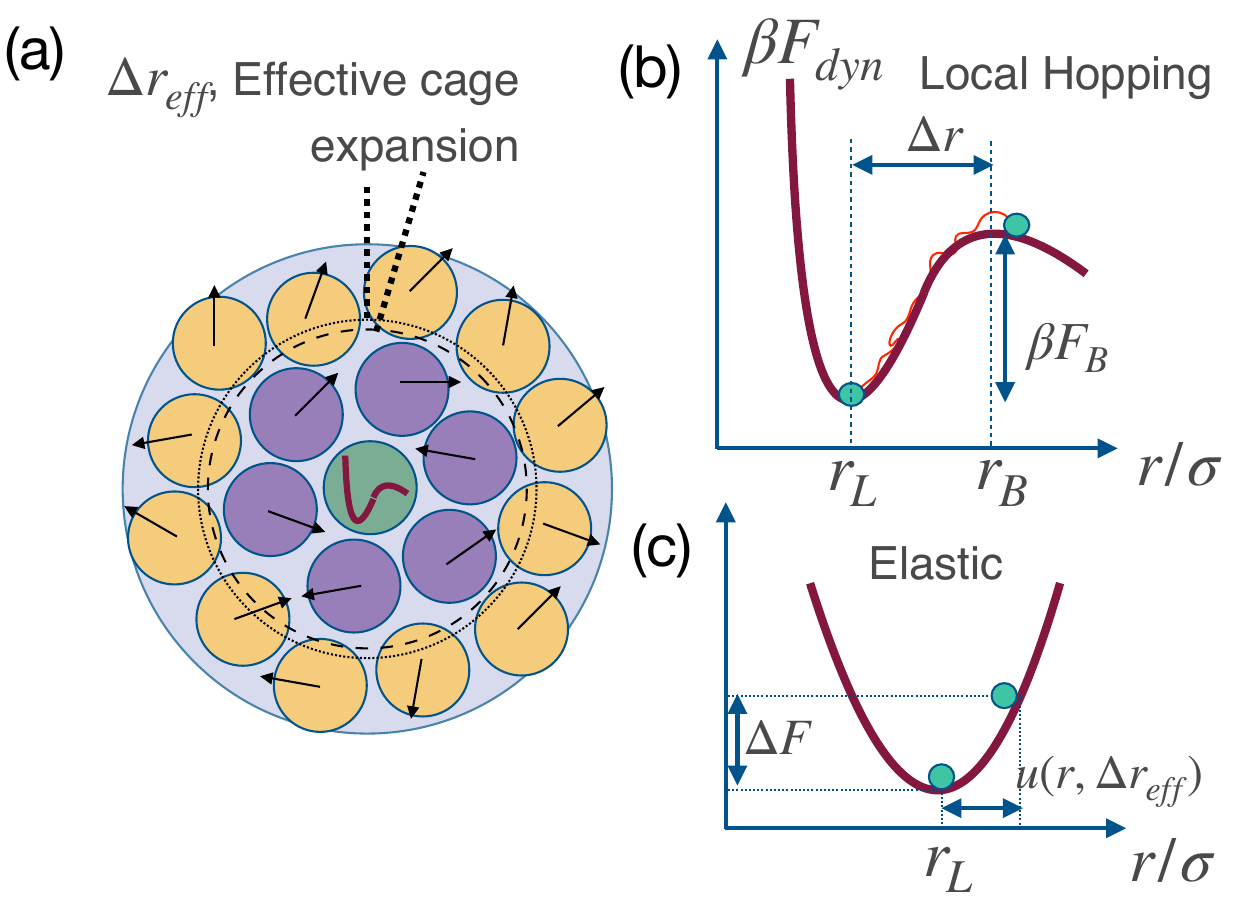}
	\caption{Schematic of the key elements of ECNLE theory \cite{Mirigian2014}: (a) Alpha or structural relaxation involves coupled cage scale hopping and longer range collective elastic dispcaments, (b) dynamic free energy as a function of scalar particle displacement with important length and energy scales indicated, (c) harmonic collective elastic displacement of particles outside the cage and the associated collective elastic barrier. 
    } 
 \label{fig2}
\end{figure}
The starting point for describing quiescent fluid activated dynamics is the overdamped Nonlinear Langevin Equation (NLE) \cite{Schweizer2005}, which is a stochastic force balance equation in for the angularly averaged scalar displacement of a tagged particle, $r(t)$: 
\begin{equation}\label{eqn:NLE} 
	-\zeta_s\frac{dr}{dt}-\frac{\partial F_{dyn}}{\partial r}+\delta f=0
\end{equation}
The first term represents the non-activated short time and distance frictional drag characterized by a well known short time friction constant $\zeta_s$, while the last term is the corresponding fluctuating white noise random force that obeys, $\left\langle\delta f\left(0\right)\delta f\left(t\right)\right\rangle=2k_BT\zeta_s\delta\left(t\right)$. This time scale associated with the dissipative non-activated short time and distance process (which could include elementary local hydrodynamic effects) ($\tau_s=\beta\zeta_sd^2$) is used as a unit of time in our analysis, and its calculation for hard and sticky spheres is well documented \cite{Ghosh2019,Schweizer2005}. {Here, $d$ is the particle diameter.}  The explicit expression for hard spheres is 
\begin{equation}
    \begin{split}
    \tau_s\equiv&\tau_0\left[1+\frac{d^3}{36\pi\phi}\int^\infty_0dq\frac{q^2(S(qd)-1)^2}{S(qd)+b(qd)} \right],\\&b^{-1}(qd)=1-j_0(qd)+2j_2(qd) 
    \end{split}
\end{equation}
Here, $j_n(x)$ is the spherical Bessel function of order $n$, and $\tau_0\equiv\frac{\zeta_{SE}d^2}{k_BT}$ is the “bare” elementary time written in terms of the Stokes-Einstein friction constant, $\zeta_{SE}$, and the contact value of the pair correlation function, $g(d)$. The modest variation of $\tau_s$ over the range of system parameters studied in this article is shown in SI Fig.S1. \\ 

The second term in Eq.\ref{eqn:NLE} is the effective force on a moving particle defined as the negative gradient of the dynamic free energy, $F_{dyn}(r)$, where $r$ is the scalar displacement of a particle from its initial position. This spatially resolved dynamic free energy is the fundamental quantity for predicting stochastic particle trajectories. For hard spheres, it encodes a particle-displacement dependent effective caging force on a tagged particle due to all the other particles (Fig.\ref{fig2}(a)), while for sticky spheres it includes both steric caging and physical bonding. An example is shown in Fig.\ref{fig2}(b), along with the relevant length and energy scales that quantify the particle transient localization and hopping. We emphasize that $F_{dyn}\left(r\right)$ is a priori calculated from knowledge of the interactions, thermodynamic state and pair structure, with the latter entering via the dimensionless static structure factor $S\left(q\right)$ and an effective force vertex, $\vec{M}\left(q\right)$, as \cite{Schweizer2005,Dell2015}, 
\begin{equation}\label{eqn:Fdyn}
    \beta F_{dyn}\left(r\right)=-3\ln{\frac{r}{d}}-\frac{\rho}{2\pi^2}\int_{0}^{\infty}{\frac{\left|\vec{M}\left(q\right)\right|^2S\left(q\right)}{1+\frac{1}{S\left(q\right)}}e^{-\frac{q^2r^2}{6}\left(1+\frac{1}{S\left(q\right)}\right)}}dq.
\end{equation} 
Here, $\beta\equiv\left(k_BT\right)^{-1}$ is the inverse thermal energy, $\rho$ the fluid number density, $\phi\equiv\frac{\pi\rho d^3}{6}$ the packing fraction, and $\left|\vec{M}\left(q\right)\right|$ is a spatially-resolved (in Fourier space) effective force vertex discussed below. The dynamic free energy first acquires the localized trapping form (Fig.\ref{fig2}(b)) at $\phi_c=0.44$ for hard spheres at the ideal näive (single particle) mode coupling theory (NMCT) transition \cite{Schweizer2005,Zhou2020} in the absence of ergodicity restoring activated hopping. The static correlations required as input to the dynamical theory are computed using highly accurate integral equation theory with the modified-Verlet closure \cite{Verlet1980,Zhou2020}.  \\

\begin{figure}
\includegraphics[width=0.45\textwidth]{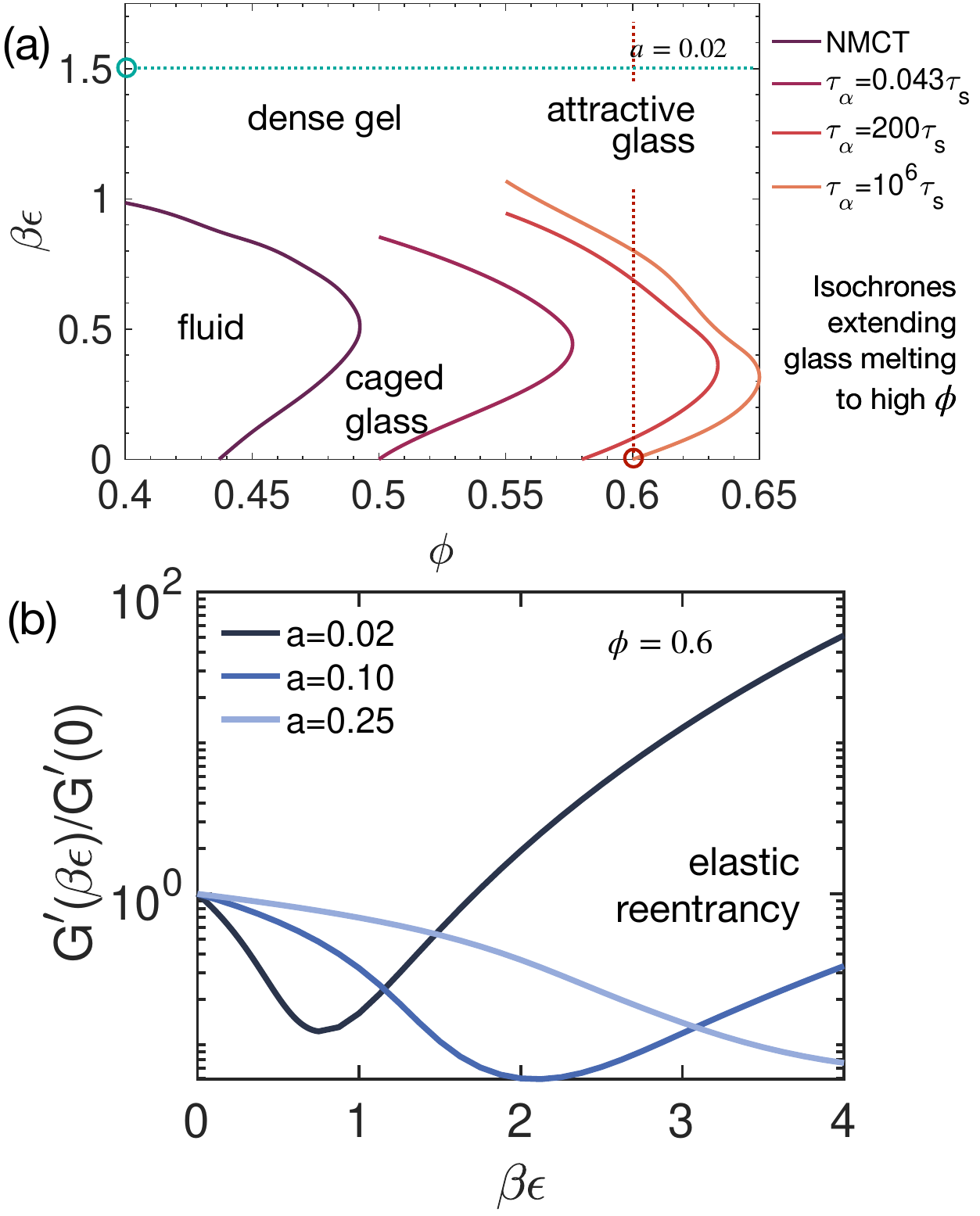}
	\caption{Attraction driven quiescent re-entrant phenomena. (a) Ideal NMCT dynamic arrest (crossover) map and analogous isochronal boundaries based on laboratory time scale activated relaxation predicted by the hybrid-PDT version of ECNLE theory \cite{Mutneja2024}. The different physical states are indicated, with the vertical dotted line at $\phi=0.6$ and the horizontal cyan dotted line at $\beta\epsilon=1.5$ indicating a specific AG state studied in detail. (b) Theory prediction of the re-entrant elastic shear modulus as a function of attraction strength for various attraction ranges \cite{Mutneja2024}. 
    } 
 \label{fig3}
\end{figure}

For $\vec{M}\left(q\right)$, we employ the relatively new hybrid-projectionless dynamic (hybrid-PDT) approach. It \textit{explicitly} treats the attractive forces in real space and has been shown to provide quantitatively and qualitatively superior results in dense quiescent fluids \cite{Dell2015,Ghosh2019,GhoshPRE2020,Mutneja2024}. This choice qualitatively differs from employing the standard MCT choice to simultaneously project both the repulsive and attractive forces on static pair density fluctuations. The hybrid-PDT effective force vertex is given by, \\
\begin{equation}\label{eqn:HybridForceVertex}
%	\begin{split}
		\vec{M}(q)=qC_0(q)\hat{q}+4\pi\hat{r}\int_0^\infty r^2f(r)g(r)\frac{sin(qr)}{qr}dr
%	\end{split}
\end{equation}
The first term accounts for the repulsive HS interaction in the usual manner (projection approximation) with $C_0\left(q\right)$ the pure HS fluid direct correlation function. The second term explicitly captures attractive forces, where $g(r)$ is the interparticle pair correlation function and $f\left(r\right)$ the attractive force. The latter is the negative gradient of the pair potential beyond contact, here taken to be of an exponential form, \begin{equation}\label{eqn:Potential}
	V(r)= 
	\begin{cases}
		\infty,& r\leq d\\
		-\epsilon e^{-\frac{r-d}{a}},& r>d
	\end{cases}
\end{equation} 
The positive parameters $\epsilon$ and a quantifying the strength and range parameters in units of $\beta^{-1}=k_BT$ and $d$, respectively. A negative interference or cross term between short range attractive and repulsive forces enters $\left|\vec{M}\left(q\right)\right|^2$ which has been shown \cite{Mutneja2024} to be critical for properly predicting re-entrant glass-melting phenomena under the ultra-dense conditions relevant to experiment. This includes a non-monotonic relaxation time and kinetic arrest boundary (Fig.\ref{fig3}(a)) \textit{and} a non-monotonic elastic shear modulus (Fig.\ref{fig3}(b)) as a function of attraction strength. The theoretical results are consistent with experiments \cite{Pham2008} and simulations \cite{Fullerton2020}.\\

The ensemble-averaged dynamic (transient) localization length in NMCT follows from the minimum of the dynamic free energy \cite{Schweizer2005} and obeys the self-consistent equation:
\begin{equation}\label{eqn:localization_length}
	\frac{1}{r_L^2}=\frac{1}{9}\int\frac{\vec{dq}}{(2\pi)^3}\rho |\vec{M}(q)|^2 S(q) e^{-\frac{q^2r_L^2}{6}[1+S^{-1}(q)]}.
\end{equation}
Based on the standard idea that slow density fluctuations control slow stress fluctuations \cite{Ngele1998}, the elastic shear modulus at the NMCT level with the hybrid force vertex is \cite{Mutneja2024}:
\begin{equation}{\label{eqn:Gprime}}
	G^\prime=\frac{k_BT}{60 \pi^2}\int_0^\infty dq \left[q^2 \rho S(q) \frac{d (|\vec{M}(q)|/q)}{dq}\right]^2\exp\left(-\frac{q^2r_L^2}{3S(q)}\right)
\end{equation}
The theory properly predicts \cite{Mutneja2024} for dense sticky colloids that the elastic modulus exhibits a non-monotonic re-entrant behavior (Fig.\ref{fig3}(b)) with increasing attraction strength, which will show below plays a very important role in understanding double yielding. This behavior is not obtained based on using the standard MCT projection approximation \cite{Mutneja2024}.\\

Activated single particle relaxation is described based on the Elastically Collective NLE (ECNLE) theory \cite{Mirigian2014}. The activation barrier involves a local cage contribution $F_B$ (discussed above), and a collective elastic component $F_{el}$ associated with a small cage expansion required to accommodate a particle hop. These aspects are visually illustrated in Figs. \ref{fig2}(b) and \ref{fig2}(c). The latter involves (i) the microscopic particle jump distance, which sets the amplitude of the elastic displacement field outside the cage, and (ii) the local material rigidity, which together determines the energy cost for harmonic displacements. The elastic barrier is calculated within the Einstein glass framework as $F_{el}=4\pi\int_{r_{cage}}^{\infty}r^2\rho g\left(r\right)\left(\frac{1}{2}K_0u\left(r\right)^2\right)dr$, where $K_0\left(r\right)$ is the harmonic spring constant of $F_{dyn}$ at its minima, and $u\left(r\right)$ is the displacement field required for a cage escape with r the distance from the cage center. The elastic displacements are predicted to be small, of order the dynamic localization length or less. Hence, the displacement field is constructed in the spirit of continuum linear elasticity \cite{Dyre2006} as $u\left(r\right)=\Delta r_{eff}\left(\frac{r_{cage}}{r}\right)^2 $for $r\geq r_{cage}$, where the cage radius $r_{cage}$ is identified as the distance at the first minimum of $g\left(r\right)$. The amplitude ($\Delta r_{eff}$), or effective cage expansion, follows from a microscopic analysis of the mean extent to which cage scale hopping results in a particle displacement larger than the cage size. Defining the microscopic jump distance $\Delta r=r_B-r_L$, this analysis yields \cite{GhoshJCP2020}
\begin{equation}\label{eqn:ECNLE_cageExpansion}
	\Delta r_{eff}\approx\frac{3}{r^3_{cage}}\left(\frac{r^2_{cage}\Delta r^2}{32} -\frac{r_{cage}\Delta r^3}{192} +\frac{\Delta r^4}{3072} \right)
\end{equation}\\

Given the elastic and local barriers, the experimentally relevant total activation barrier is $F_{total}=F_B+F_{el}$. The mean structural or alpha relaxation time follows from the Kramers expression for barrier crossing as \cite{Mirigian2014,Schweizer2005,Kramers1940}
\begin{equation}\label{eqn:TauAlpha}
	\frac{\tau_\alpha}{\tau_s}=e^{\beta F_{el}} \int_{r_L}^{r_B}dx~e^{\beta F_{dyn}(x)}\int_{r_L}^x dy~e^{-\beta F_{dyn}(y)}
\end{equation}\\
For a total barrier beyond the low value of $\sim 1-2~k_BT$, the above expression reduces to \cite{Mirigian2014,Schweizer2005} $\tau_\alpha/\tau_s\approx\left(2\pi/\sqrt{K_0K_B}\right)e^{\beta F_{total}}$. Here, per Figs.\ref{fig2}(b) and \ref{fig2}(c),  $K_0$ and $K_B$ are the absolute values of the curvatures of dynamic free energy at its minimum (the localization length $r_L$) and maximum (barrier location $r_B$), respectively. Using the Kramers time, the glass-melting kinetic arrest boundary can be extended to the very high packing fraction regime well beyond the NMCT crossover, which is relevant to experiments and simulations based on alpha time isochrones. An example result is shown in Fig.\ref{fig3}(a) and is in good accord with experiment and simulation \cite{Mutneja2024}.\\

 \subsection{Rheological Framework and External Stress} 
We adopt the previously developed generalized Maxwell constitutive equation description for a system sheared at a fixed rate (accumulated strain $\gamma=\dot{\gamma}t$), which in an integration through transients form is given by \cite{Chen2011, GhoshJOR2023}
\begin{equation}\label{eqn:Constitutive}
\sigma\left(\gamma\right)=\int_{0}^{\gamma}{d\gamma^\prime G^\prime\left[\sigma\left(\gamma^\prime\right),\gamma^\prime\right]exp\left(-\int_{\gamma^\prime}^{\gamma}{d\gamma^{\prime\prime}\frac{1}{\dot{\gamma}\tau_\alpha\left[\sigma\left(\gamma^{\prime\prime}\right),\gamma^{\prime\prime}\right]}}\right)}
\end{equation}
or in differential form by, $\frac{d\sigma\left(\gamma\right)}{d\gamma}+\frac{\sigma\left(\gamma\right)}{\dot{\gamma}\tau_\alpha\left[\sigma\left(\gamma\right),\gamma\right]}=G^\prime\left[\sigma\left(\gamma\right),\gamma\right]$. This is a self-consistent equation since stress depends on the nonlinear elastic modulus $G^\prime\left(\sigma,\gamma\right)$ and the structural (stress) relaxation time $\tau_\alpha\left(\sigma,\gamma\right)$, and vice-versa. Per above, recall the relaxation time is expressed in units of  $\tau_s$, which enters only as a prefactor in the alpha time and is taken to be invariant to deformation. Deformation leads to elastic stress build-up via $G^\prime$, while stress can be dynamically relaxed via structural relaxation on a timescale $\tau_\alpha$. An explicit tensorial description is not adopted. This is similar to other theories in the literature \cite{Falk1998,FuchsFarday2002}, including the recent successful theory \cite{Ghosh2023} for stress overshoots and {step rate} rheology of dense hard sphere colloidal suspensions. Importantly, the effectively-isotropic on the rheologically relevant length scale simplification is consistent with the scalar displacement-based ECNLE theory for particle motion \cite{Mirigian2014,Schweizer2005}, and also the need to retain tractability since the microscopic pair structural input is required. Simulations of the rheology of glassy fluids have also provided significant evidence that anisotropic effects on the microscopic scales relevant to our approach are small \cite{Miyazaki2004,Yamamoto1998}.\\

We also consider two simplified rheological protocols since they are relevant to understanding our new predictions for the {step rate} rheology of dense sticky suspensions and are interest for their own sake. The first relates to a step-strain experiment where a system is instantaneously strained by an amount $\gamma$ at time $t=0$. The induced instantaneous stress ($\sigma_{step}(t=0^+)$) is then measured to obtain the $\sigma_{step}-\gamma$ curve, along with the complete nonlinear stress-relaxation response. Eq(\ref{eqn:Constitutive}) then reduces to, 
\begin{equation}
{\sigma}_{step}\left(\gamma,t\right)=G^\prime\left(\sigma_{step},\gamma\right)\gamma\ e^\frac{-t}{\tau_\alpha\left[\sigma_{step}\left(\gamma,t\right),\gamma\left(t\right)\right]}
\end{equation}
Our focus here is only the $\sigma_{step}-\gamma$ response at $t=0^+$ for which:
\begin{equation}\label{eq:StepStrain}
	\sigma_{step}(\gamma,t=0^+)=G^\prime(\sigma_{step},\gamma)\gamma
\end{equation}\\
The second limiting scenario is to ignore entirely thermally-driven dynamical relaxation in the spirit of an infinite alpha time corresponding to a nonlinear elastic scenario. This limit will be instructive in understanding our full {step rate} startup shear rheology results and is also motivated by the quasi-static shear response of granular matter \cite{FarainPRL2023} and, more generally, as a limiting model for Brownian systems. Taking the $\tau_\alpha\rightarrow\infty$ limit in Eq(\ref{eqn:Constitutive}) yields: 
 \begin{equation}\label{eq:TauInftyStrain}
	\sigma(\gamma)=\int_0^\gamma d\gamma' G^\prime[\sigma(\gamma'),\gamma']
\end{equation}\\

Implementation of Eqs. [\ref{eqn:Constitutive}], [\ref{eq:StepStrain}], and [\ref{eq:TauInftyStrain}] require \textit{self-consistent} determination of the effect of stress, shear rate, and strain on the elastic modulus and alpha time. Prior ECNLE theory-based rheology work \cite{Kobelev2005,GhoshJCP2020} adopted the heuristic, but argued to be qualitatively and physically sound, inclusion of external stress via its transduction down to a microscopic force $f_{ext}$ on a tagged particle in the spirit of microrheology. Deformation-induced changes on the relevant local structural scale were ignored. The theoretical results properly captured important experimental phenomena like deformation-induced mobility enhancement, inverse power-law shear thinning of the alpha time and viscosity, a Hershel-Buckley flow curve, and an exponential growth with packing fraction of the steady-state stress \cite{Ghosh2019}. Specifically, the external stress enters as a mechanical work term in the \textit{nonequilibrium} dynamic free energy as, 
\begin{equation}\label{eqn:StressAddition}
    \beta F_{dyn}\left(r,\sigma\right)=\beta F_{dyn}\left(r,\sigma=0\right)-f_{ext}r,
\end{equation}
where macroscopic stress is related to the microscopic force via the particle cross sectional area, $f_{ext}=A\sigma$, with $A=\frac{\pi d^2}{24}$ \cite{Kobelev2005,Ghosh2019}. As shown in ref \cite{Ghosh2019}, different choices of $A$ only modestly change the absolute values of the key quantities, not qualitative features.\\
\begin{figure}
	\includegraphics[width=0.45\textwidth]{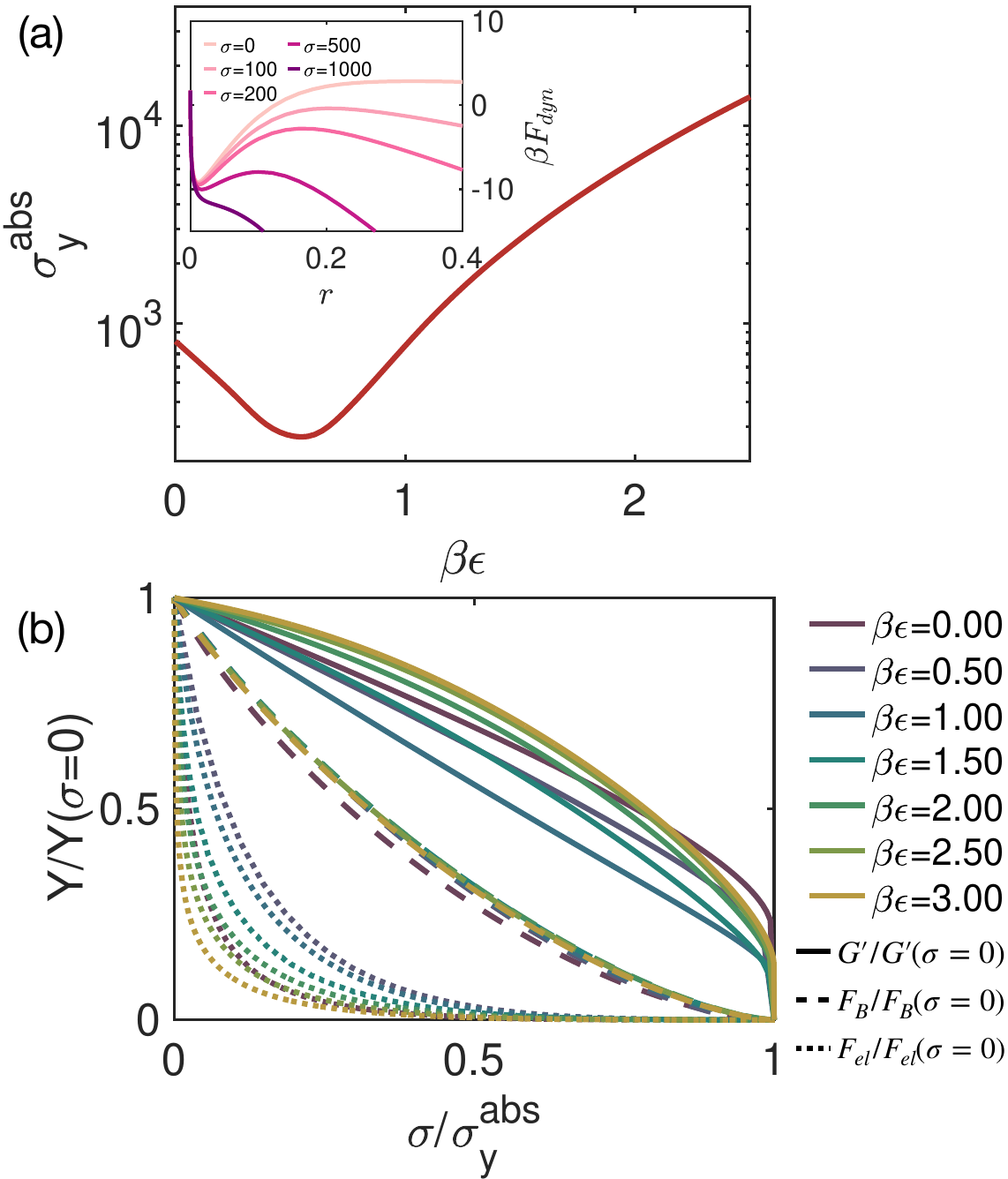}
	\caption{ Effect of external stress based on a deformation-invariant $S(q)$. (a) Inset: example of how the dynamic free energy becomes less localized with increasing stress for an attractive glass with $\phi=0.60$, $\beta\epsilon=1.5$, and $a=0.02$. Main panel: corresponding variation with attraction strength of the absolute yield stress.  (b) Doubly normalized plot of the local cage (dashed) and collective elastic (dotted) barriers in thermal energy units as a function of stress for $\phi=0.60$ and various attraction strengths and a range $a=0.02$. The solid curves are the corresponding results for the elastic shear modulus. The quiescent state local and elastic barriers for the shown attraction strengths are [$12.0,~3.9,~27.7,~107.6,~273.5,~565.4,~1034.9 $] and [$8.8,~0.4,~22.0,~203.3,~1018.9,~2682.6,~12519.0$], respectively. 
 } 
	\label{fig4}
\end{figure}
External stress (microscopic force) thus reduces the activation barrier (both the local cage and collective elastic) and elastic modulus. An example of the change in the dynamic free energy and barrier reduction is shown in the inset of Fig.\ref{fig4}(a). At a sufficiently high stress, a delocalization or fluidization transition is predicted whereby the minimum of the dynamic free energy is destroyed at a critical “absolute yield stress”, $\sigma_y^{abs}$. Results for this quantity are shown in the main panel of Fig.\ref{fig4}(a) for attractive sphere suspensions at $\phi=0.60$ with different attraction strengths. The predicted non-monotonic trend is consistent with quiescent glass melting behavior at high packing fractions (see Fig.\ref{fig3}(b)). Fig.\ref{fig4}(b) shows the normalized stress evolution of the local and elastic barriers for different attractive systems. Apart from the known trend that elastic barriers are more rapidly reduced with stress than their local analog \cite{GhoshJCP2020}, the overall response is qualitatively independent of attraction strength. The same is true for the stress evolution of the elastic modulus in Fig.\ref{fig4}(b). Moreover, since stress is assumed to not change packing correlations and hence not the kinetic constraints, we do not expect much variation of the normalized elastic modulus or yield stress per Fig.\ref{fig4}(b).\\

\section{Theory Generalization: Deformation-Induced Structure Changes, Activated   Relaxation, and Elementary Rheological Consequences}\label{Sec3}
The above approach does not address how deformation changes the key structural correlations that quantify kinetic constraints and the elastic and relaxational properties. This isostructural simplification has been proven to \textit{not} predict an overshoot in the transient stress-strain curve in the generalized Maxwell model framework \cite{GhoshJOR2023}. Here, we propose a new, fully self-contained, self-consistent approach to this problem, which forms the basis of all our results. The first step requires a proposal for how deformation modifies structure. For Brownian suspensions \textit{not} close to a shear thickening or jamming regime \cite{BonnRevModPhys2017}, the local caging correlations, and consequently the kinetic constraints, weaken on average with shear-induced structural deformations. This effect is expected to become even more significant and rich for systems with competing forces and length scales, such as attractive glasses. The second step generalizes ideas successfully employed for the rheology of polymer glasses \cite{Chen2011} and entangled polymer fluids \cite{Sussman2012} for the nonequilibrium temporal evolution of structure and driving force for re-equilibration.\\

\subsection{Strain-induced structural softening} 
\begin{figure}
	\includegraphics[width=0.4\textwidth]{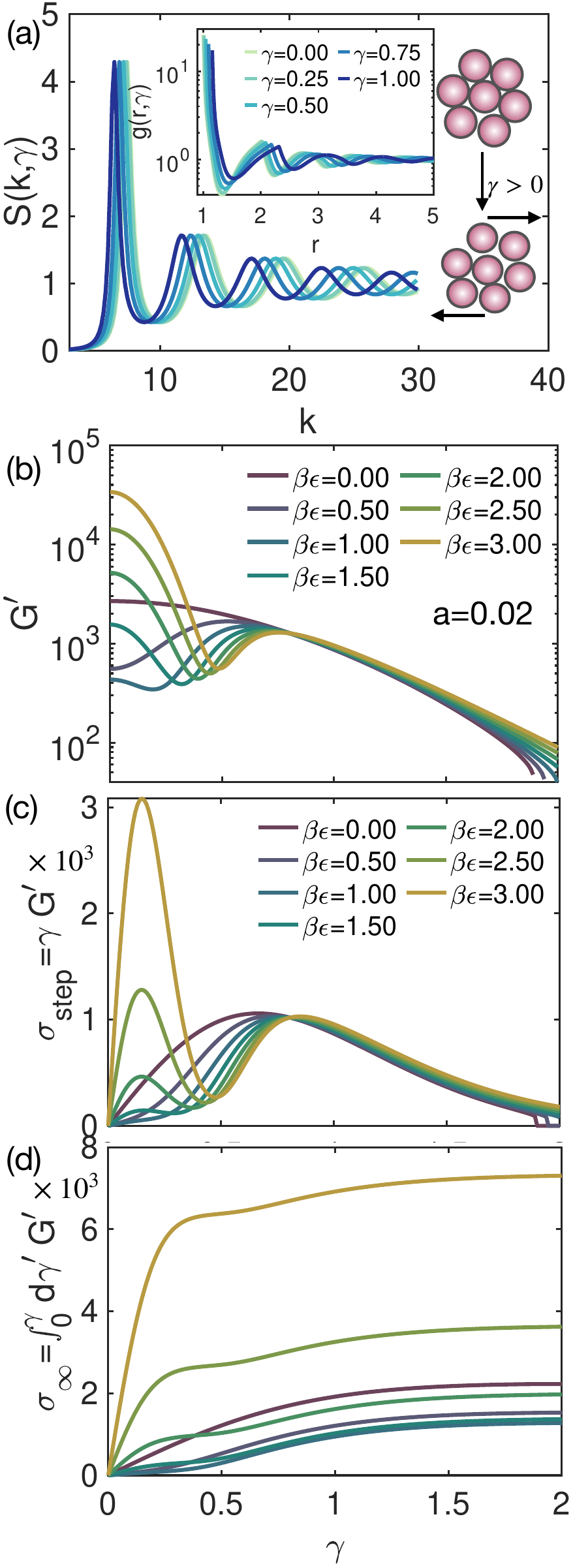}
	\caption{\textbf{Effects of strain induced deformation:} (a) Evolution of the structure factor under imposed strain for an attractive glass with $\phi=0.60$ and attraction strength and range of $\beta\epsilon=1.5$ and $a=0.02$, respectively. The inset depicts the corresponding evolution of the pair-correlation function and strain-induced cage expansion. (b) Strain-induced evolution of the elastic modulus for different attraction strengths at a fixed range and $\phi=0.60$. (c) Stress-strain curves for an idealized instantaneous step strain experiment immediately after deformation for systems with the same attraction strengths as panel (b). Note the predicted double overshoot behavior. (d) Stress-strain curve predicted by the new theory in the limit of no dynamical relaxation.} 
	\label{fig5}
\end{figure}
We adopt a minimalist, \textit{no} adjustable parameter, physically-motivated version of the isotropic wave-vector “shear advection” idea proposed in the context of the ideal MCT of colloid rheology \cite{Fuchs2010,Amann2013,AmannJOR2014,Fuchs2002,FuchsFarday2002,Fuchs2005}. In that approach,  structural correlations and dynamic time correlations enter a generalized stress-stress memory function where flow-induced changes of quiescent quantities in Fourier space are modified via a strain-induced advected wavevector $k(t)\rightarrow k\sqrt{(1+\left(\dot{\gamma}t/\gamma_c\right)^2)}$, where $\gamma_c$ is an adjustable parameter. Advection reduces caging constraints as a function of accumulated strain $\gamma=\dot{\gamma}t$. However, as recently discussed \cite{GhoshJOR2023,Amann2013}, the ideal MCT does not capture the experimentally observed non-monotonic evolution of the overshoot magnitude with strain rate and the vanishing overshoot amplitude at ultra high densities for HS suspensions. To our knowledge, it also does not capture the rich double yielding phenomena nor the non-monotonic evolution of the equilibrium elastic shear modulus in attractive glasses. But we do believe its treatment of how the structure is modified under deformation via advection is physically sound as a starting point and adopt it. Strain-induced wavevector advection of the structure factor maintains its quiescent functional form ($S_0\left(k\right)$) and corresponds to:  $S\left(k,\gamma\right)=S_0\left(k\sqrt{\left(1+\gamma^2/3\right)}\right)$. The corresponding real space pair correlation function is $g\left(r,\gamma\right)=1+\frac{g_0\left(r/\sqrt{1+\gamma^2/3}\right)-1.0}{\left(1+\gamma^2/3\right)^\frac{3}{2}}$. Example strain evolutions of the structure are shown in Fig.\ref{fig5}(a).\\

Now, for a system with an attraction range “$a$”, { which is well defined for the exponential model potential we adopt,} the amount of strain required to reduce the dimensionless bond strength, $\beta\epsilon$, by a factor of $1/e$ is simply $\sqrt{3\left(\left(1+a\right)^2-1\right)}$. For $a=0.02$, this yields $\gamma=0.348$. Of course, at this strain the absolute magnitude of attraction is not necessarily negligible. However, when it becomes comparable to the strength of the underlying caging constraints of hard spheres, the negative interference term in Eqs. (\ref{eqn:Fdyn}) and (\ref{eqn:Gprime}) via the effective force vertex will begin to become important, thereby reducing the net total force constraints on particle motion \textit{below} that of pure hard spheres. With further straining, bonds weaken so much that they can be viewed as “broken”, and the system reverts to a HS-like system.\\

The above strain evolution is effectively captured by the elastic modulus results in Fig.\ref{fig5}(b) for different attraction strengths. For attractive glasses, the initially high elastic modulus due to the presence of strong bonding and caging quickly decreases with strain due to bond weakening and drops to a value below that of hard spheres due to this force interference effect. Eventually, the shear modulus goes through a minimum and then rises to the HS value at sufficiently large strains. These results support and quantify the proposed intuitive picture \cite{Pham2008,Koumakis2011,Moghimi2020} that at high enough strain, physical bonds are “broken”, and at higher strains the material from an elastic perspective is akin to a HS system. Of course, the strain required to completely eliminate the bonding effect increases monotonically with attraction strength.The evolution of the local and elastic activation barriers and mean alpha relaxation time follow a similar pattern, for the same physical reasons presented above (see \textit{SI} Fig.S3). \\

The predicted non-monotonic evolution of the elastic modulus with structural deformation (strain) can be used to obtain the step-strain response. Per Eq(\ref{eq:StepStrain}), the product $G^\prime\gamma$ defines the instantaneous stress. Results are shown in Fig.\ref{fig5}(c) for an attractive glass with different attraction strengths of fixed $a=0.02$ at $\phi=0.60$. A double yielding like behavior is predicted, per the 2-step elastic modulus decay in Fig.\ref{fig5}(b), {of entirely nonlinear elastic origin.} The first overshoot thus represents strain-induced bond breakage, while the second overshoot indicates the breakup of perturbed (much more ductile) caging constraints. We argue below this finding is the \textit{core mechanistic physical origin} of double yielding, although it is not the entire story for transient rheology since the present analysis is only for the nonlinear elastic response. { Quantitative application of our theory to experimental measurements performed using this protocol are given in section \ref{Sec5A}.} \\

If we ignore thermal fluctuation-driven activated stress relaxation during a startup continuous shear experiment by assuming $\tau_\alpha=\infty$ in Eq(\ref{eqn:Constitutive}), the results in Fig.\ref{fig5}(d) are predicted. An initial linear stress growth is followed by a quasi-plateau-like feature, which is then followed at larger strains by a \textit{second} even weaker stress increase for attractive glasses (per a “second elastic modulus”), leading to a final accumulated stress that mimics a nonequilibrium flow stress plateau. Hence, even without taking into account how deformation massively speeds up relaxation, a type of double yielding response is predicted. However, there is no shear rate in these calculations, akin to quasi-static mechanical tests of granular materials \cite{FarainPRL2023,BonnRevModPhys2017,Berthier2024}. Though the theory implemented at this simplified rheological level does predict a 2-step yielding, its strain dependence is of a more “plastic-like” nature with no overshoots. More generally, this two-step plastic-like form suggests the importance of deformation-modified relaxation processes in determining the striking stress overshoots, the coordinates of which are experimentally observed \cite{Pham2008,Koumakis2011,Moghimi2020} to depend significantly on shear rate and other variables.\\

\subsection{Relaxation Induced Structural Rejuvenation} 
For the Brownian systems of interest, relaxation-induced constraint built-up is necessary to reach a steady state. In the recent theoretical study of HS suspensions of Ghosh and Schweizer \cite{ GhoshJOR2023}, the structural evolution with deformation (strain) was halted by hand at a certain strain value, $\gamma_c$, based on the physically-motivated \cite{ GhoshJOR2023} criterion (which has experimental support \cite{ GhoshJOR2023, Bennin2019}) that the renormalized Peclet number $Pe=\dot{\gamma}\tau_{\alpha\ }\left(\gamma,\dot{\gamma}\right)$ reaches a constant value of $Pe\left(\gamma_c\right)\in\left(~\ 0.1-0.3\right)$. This theory made accurate predictions for the non-monotonic change of the overshoot magnitude with strain rate and functional form of the flow curve.\\

Here, we propose a new, self-consistent, fully self-contained theory for nonequilibrium structural relaxation and equilibration. We are inspired by a successful idea developed for predicting the nonlinear mechanical response of deformed polymer glasses \cite{Chen2011} where the deformation distorted structure factor $S^k\left(\gamma\right)$ ($k$ is a wavevector) relaxes towards the equilibrium structure ($S^k\left(\gamma=0\right)$) on a timescale of $\tau_\alpha\left(\gamma,\sigma,\dot{\gamma}\right)$. The key physical idea is the intuitive notion of an “effective strain”, $\gamma_{eff}$, that quantifies in a simple and tractable scalar manner the degree of structural distortion as a function of $\gamma$. It always initially grows linearly with strain (or elapsed time in a startup deformation) in an elastic solid-like regime. But at long times in a nonequilibrium flow steady state, it saturates, marking the cessation of structural evolution and onset of viscous flow. The evolution equation for the structure factor is proposed to be:
\begin{equation}\label{eqn:SkEvolution}
\begin{split}
     \frac{dS^k\left(\gamma\right)}{d\gamma}=&-f\frac{S^k\left(\gamma\right)-S^k\left(0\right)}{\dot{\gamma}\tau_\alpha\left(\gamma,\sigma,\dot{\gamma}\right)\ }\\&{+\left(\frac{k\gamma}{3\sqrt{\left(\gamma^2/3+1\right)}}\left.\frac{\partial S^k\left(\gamma\right)}{\partial k}\right|_{k=k\sqrt{1+\gamma^2/3}}\right)}_{\gamma=\gamma_{eff}}
\end{split}
\end{equation}
where notationally $S^k\left(\gamma\right)\equiv S\left(k;\gamma\right)$. The first term is the driving force for structural relaxation towards the undeformed $\gamma=0$ state. The second term of the opposite sign is the driving force for affine structural change which reduces kinetic constraints based on the wavevector advection idea and $\gamma_{eff}$. The prefactor $f$ is a possibly nonuniversal parameter that quantifies the connection between the structural relaxation time and the mean alpha or stress relaxation time in the constitutive Eq.(\ref{eqn:Constitutive}). If the latter two processes are slaved, then $f=1$, which we believe is a reasonable minimalist ansatz, and is adopted in all the calculations presented below. The effect of this parameter is studied in the \textit{SI} (Fig.S4) and appears to have no qualitative consequences on our results.\\

The structural evolution per Eq.(\ref{eqn:SkEvolution}) is complicated since it involves all wavevectors. To simplify in a manner consistent with how we model structural changes via wavevector advection, we make what we view as a natural and physically plausible assumption that the structure recovers in the same manner as it was distorted, i.e., with the functional form intact and shifting of the wavevector (advection). This tremendously simplifies the description of nonequilibrium structural evolution. One can then rewrite Eq.(\ref{eqn:SkEvolution}) solely in terms of  a shifted wavevector description:
\begin{equation}\label{eqn:kEvolution}
\frac{dk^\prime\left(\gamma\right)}{d\gamma}=-f\frac{k^\prime\left(\gamma\right)-k^\prime\left(0\right)}{\dot{\gamma}\tau_\alpha}-\left(\frac{k^\prime\left(0\right)\gamma}{3\left(\gamma^2/3+1\right)^{1.5}}\right)_{\gamma=\gamma_{eff}}
\end{equation}

with initial conditions $k^\prime\left(\gamma=0\right)=k$ and $S^k\left(\gamma\right)\equiv S^{k^\prime}$. The effective strain that quantifies the net structural deformation follows as: $\gamma_{eff}=\sqrt3\left[\left(k^\prime\left(0\right)/k^\prime\left(\gamma\right)\right)^2-1.0\right]^{1/2}$. As opposed to the accumulated strain $\gamma$ per an affine elastic deformation, $\gamma_{eff}$ evolves non-linearly in time and approaches a nonequilibrium steady state limit that signals the attainment of a nonequilibrium steady state $S\left(k\right)$. \\

To summarize, Eqs. (\ref{eqn:Constitutive}) and (\ref{eqn:kEvolution}), in combination with the wavevector advection idea and ECNLE theory for the elastic modulus and relaxation time under deformation, form a set of closed equations for the coupled stress and structural nonequilibrium dynamics. These equations can be numerically solved self-consistently to obtain the stress-strain curve. Below, we apply this framework to dense attractive particle systems to broadly explore the double yielding and other phenomena. We note that for the HS system, the predictions of the present theory qualitatively align well with those presented in Ref. [40], including the behavior of the stress overshoot. The HS prediction is illustrated in our figures as the $\beta\epsilon=0$ special case per Fig.\ref{fig6}(a), and discussed further in \textit{SI} Fig.S2. We emphasize that in qualitative contrast to the approach in Ref. \cite{GhoshJOR2023}, we do not impose a predefined cutoff strain for structural evolution and nonequilibrium equilibration. Rather, the latter is a priori predicted, and a nonequilibrium steady state emerges smoothly.\\

\section{Step rate Rheology: Hard Spheres to Attractive Glasses to Dense Gels}\label{Sec4}
We now present representative theoretical predictions for the nonlinear {step rate} rheology focusing on the transient regime and its \textit{qualitative} changes as the system evolves from HS colloids to attractive glasses and dense gels as a function of packing fraction, attraction strength and spatial range, and shear rate. A prime result is the correct theoretical prediction and a microscopic physical understanding of the evolution of the mechanical response from pure ideal plastic like flow (no overshoot) to a single overshoot response associated with tight caging, to a double yielding response indicative of a competition of physical bonding and ductile caging. As observed experimentally \cite{Pham2008,Koumakis2011,Moghimi2020}, this evolution is predicted to be realizable based on increasing particle concentration, attraction strength, and shear rate, or reducing the attraction spatial range.  \\

\subsection{Role of Attractive Potential Strength}
\begin{figure}
	\includegraphics[width=0.47\textwidth]{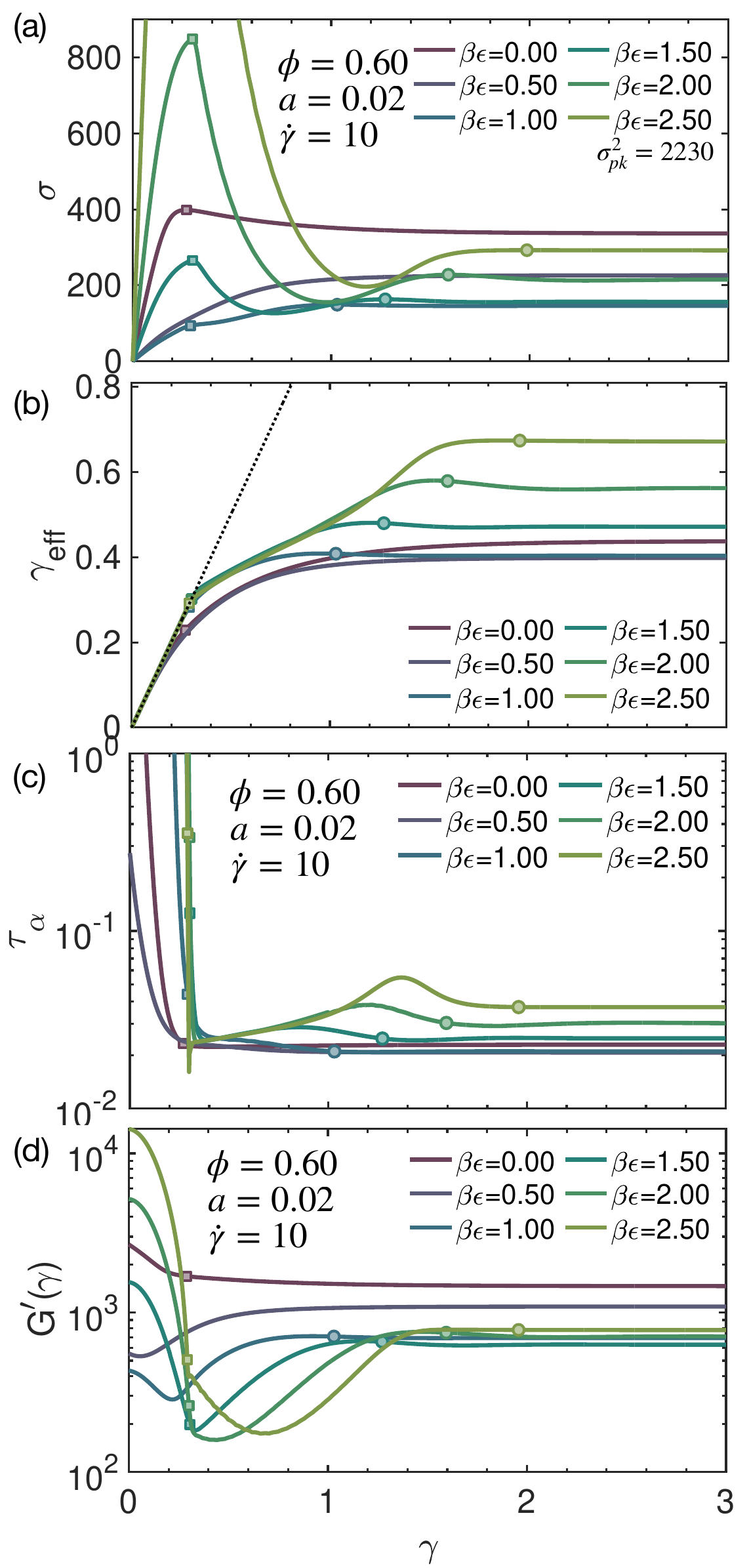}
	\caption{ {Step rate} shear rheology of attractive glasses: Effect of bond strength. An attractive glass at $\phi=0.6$ with a fixed attraction range of $0.02$ and dimensionless shear rate of 10 as a function of attraction strength. (a) Dimensionless stress-strain curves. The $\beta\epsilon=0$ curve is the reference HS system. The off-scale peak for $\beta\epsilon=2.5$ has coordinates of $\left[0.3,\ 2230\right]$. (b) Evolution of the deformation-induced structural softening metric ($\gamma_{eff}$) which initially grows linearly with strain, then exhibits sublinear softening, and eventually saturates in the steady state. A weak overshoot is predicted for intermediate attraction strengths. (c) and (d) quantification of deformation softening of the relaxation time and elastic modulus, respectively. Their steady state values both show a non-monotonic, but quantitatively different, behavior with attraction strength, similar to the quiescent fluid. The steady state dimensionless Peclet number is $Pe\in\left[0.2-0.4\right]$. Quiescent mean alpha times with increasing attraction strength are $[8.76,\ 0.345,\ 22.04,\ 203.40,\ 1017,\ 3683]$.}
	\label{fig6}
\end{figure}
Figure \ref{fig6}(a) shows the stress-strain curve evolution as a function of attraction strength, the vertical trajectory in Fig.\ref{fig3}(a). The $\beta\epsilon=0$ curve denotes the HS reference state which exhibits a single overshoot at $\gamma\approx0.27$. As attraction strength grows, the linear shear modulus first decreases due to glass melting and elastic re-entrancy (per Fig.\ref{fig3}(b)), with the overshoot disappearing for $\beta\epsilon=0.5$. Thus, the material yields and flows in a rapid, ideal plastic-like manner. When $\beta\epsilon=1.0$, an intriguing two-step, but still plastic-like, behavior emerges. As bonds further strengthen, a two-step stress overshoot (marked by points) double yielding response grows in. Overall, we thus predict that dense sticky suspensions can exhibit zero (of 2 types), one, or two stress overshoots depending on the tunable bond strength, qualitatively consistent with experiments \cite{Moghimi2020}. The physical mechanism underlying these 4 distinct nonlinear responses involves two competing effects: (i) strain-induced structural deformation (Fig.\ref{fig5}), which weakens and eventually “breaks” bonds, and (ii) increasing stress lowers the activation barrier, thereby accelerating structural and stress relaxation in a strain rate dependent manner. The first overshoot corresponds to bond breakage, but which is a brittle event with a low yield strain. In contrast, the second overshoot arises from repulsive cage breaking, which is far more ductile than for pure HS systems, occurring at a remarkably large yield strain of $100-200\%$ strain versus only $\sim25\%$ for hard spheres. Below we further discuss the key physics underlying double yielding. \\

With increasing attraction strength, the yield stress associated with bond breakage (first peak) increases significantly compared to the cage breakup overshoot. However, the yield strain ($\gamma_{y,1}$) is governed by the bond length (attraction range) and remains unchanged with varying $\beta\epsilon$. In contrast, the second yield peak ($\gamma_{y,2}$) associated with cage rearrangement occurs at larger strain values as bond strength grows, reaching exceptionally large values of $200\%$. This feature has long been a major experimental puzzle \cite{Pham2008,Koumakis2011,Moghimi2020,Wagner_Mewis_2021}. Clearly, the system is not akin to a literal HS fluid after bond breakage, signifying the nature of the caging constraints is \textit{qualitatively} altered in the presence of strong bonding. Our results provide the first theoretical prediction of this fascinating behavior.\\

Recall from Fig.\ref{fig5} that the strain needed for the net bond strength to reach a value of $\epsilon/e$ is $\sim0.35$ for an attraction range of $a=0.02$. As attraction strength increases, a larger deformation is required to eliminate inter-particle bonding. Additionally, once the bond strength decreases enough that the first yield point is reached, the elastic modulus has been significantly reduced to values lower than that of hard spheres due to deformation-induced glass-melting. This behavior is seen in Fig.\ref{fig6}(d)), and its analog for the massive reduction of the relaxation time in Fig.\ref{fig6}(c). The latter speed-up quickly leads to a steady state if the system is pure hard spheres. But for attractive particles, further structural deformation after the first overshoot occurs, which can increase the alpha time and elastic modulus, as shown in Fig.\ref{fig5}. This allows even further elastic structural deformation to occur and, in turn, enhanced solidification, resulting in the non-monotonic features in Figs \ref{fig6}(c) and \ref{fig6}(d), that depend on shear rate (discussed below). Eventually, this deformation-induced constraint build-up ends, and the system goes through a second yield overshoot and ultimately attains a steady state. \\

Thus, our new mechanistic insight is that the double yielding behavior and exceptionally large second yield strain are consequences of deformation-induced constraint build up occurring at larger strains due to the glass melting phenomenon during bond stretching and breaking, which are complex dynamical phenomena typically discussed under quiescent conditions. This physical behavior also implies that for \textit{extremely} short-range attractions where deformation-induced constraint buildup occurs on an ultra-small strain scale, the second yield point will eventually shift back to its much smaller HS system value, as we will explicitly show below.  \\

Figure \ref{fig6}(b) presents the evolution of the metric of scalar local structural order, $\gamma_{eff}$, with strain. The results reinforce the physical explanations given above. The HS system transitions in a relatively simple manner from linear elastic response ($\gamma_{eff}=\gamma$) to viscous steady state flow beyond the overshoot. In stark contrast, for attractive glasses, the structure deforms with the applied strain until the bonds weaken (or “yield”) at the first overshoot, after which the activated relaxational dynamics drive the system to a steady state. But further deformation leads to constraint build-up, per Fig.\ref{fig5}, and thus a weaker elastic-like deformation per the lower than unity slope of the $\gamma-\gamma_{eff}$ plot in Fig.\ref{fig6}(b). The reduced rate of deformation can be represented as linear growth of $\gamma_{eff}$ up to $0.25$, with an effective slope that depends on the strain rate (see next sub-section). Eventually, the structure attains its steady state beyond the second overshoot, which is more deformed the larger the attraction strength. This latter prediction is experimentally testable by measuring the evolution of the near-contact region of the pair correlation function in the steady state.\\

Figures \ref{fig6}(c) and (d) show the strain-softening of the relaxation time and elastic modulus, respectively. Their extremely high quiescent values for attractive glasses are quickly reduced to below that of hard spheres until the first yield point is reached. This results in faster structural and stress relaxation, but deformation-induced constraint build-up then allows the system to regain strength, thereby increasing the alpha time and elastic modulus, resulting in the non-monotonic evolution of these properties seen in the figures.   \\  

 \subsection{Role of Attraction Range} 
\begin{figure}[t]
	\includegraphics[width=0.45\textwidth]{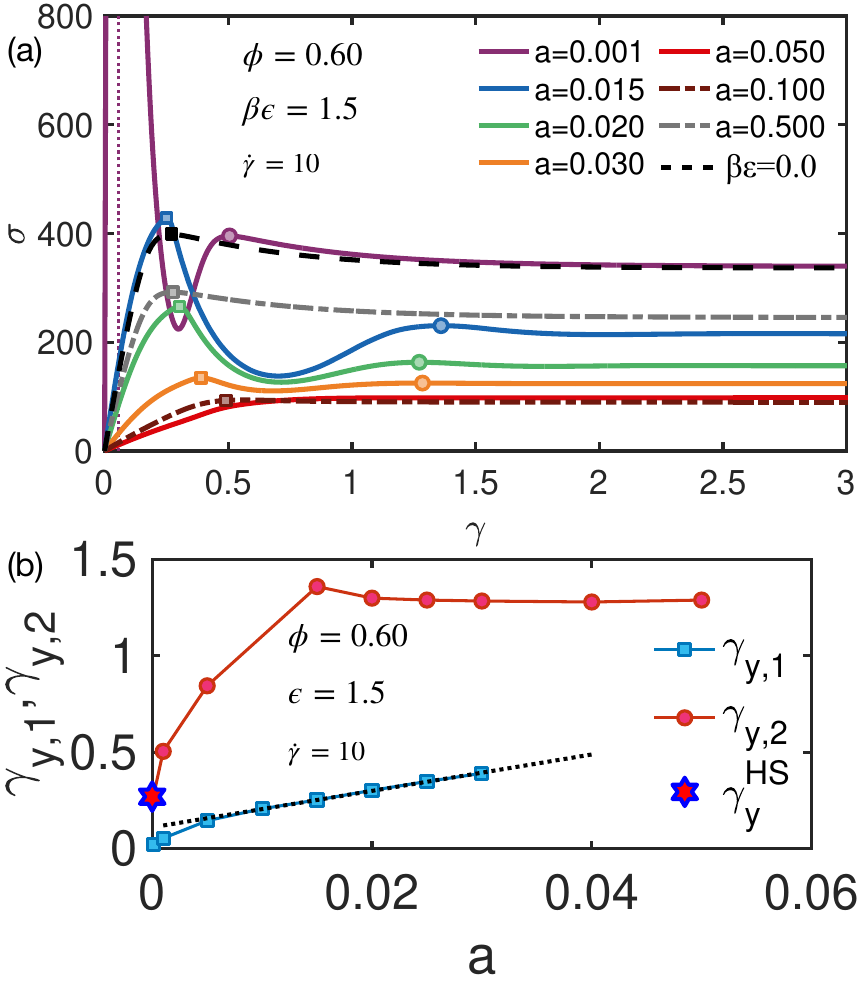}
	\caption{ {Step rate} shear rheology of attractive glasses: Role of attraction range. (a) Stress-strain curves for an attraction strength $\beta\epsilon=1.5$ and varying spatial ranges at fixed shear rate and $\phi=0.60$. {Attraction range increases from top to bottom for the solid curves . For the largest spatial range of $a=0.5$ (gray dashed-dotted curve)} the response resembles that of the HS system (black dashed line). As the range decreases, the initial modulus changes non-monotonically and a double yielding behavior emerges around $a=0.03$. The overall stress scale also varies non-monotonically with a. The first yield point associated with bond breaking shifts linearly with a (see panel (b)), while the second yield point associated with cage rearrangement approaches the HS yield strain (marked by the star symbol) as the attraction range decreases (also shown in (b)). For a very short range attraction of $a=0.001$, the bond peak stress is $\sigma_{pk}^1=5350$, with the strain level indicated by a vertical dotted line. After the second yield point, the response increasingly resembles that of a hard sphere system, a trend that becomes more pronounced with a lower attraction range and is evident from $a=0.001$ plot, while even lower ranges are detailed in the \textit{SI}.}
 \label{fig7}
\end{figure}

The attraction range sets the elementary length scale of physical bonds and the attractive \textit{force} scales as -$\frac{\epsilon}{a}$. Thus, the attraction range is expected to have \textit{qualitative} consequences for key features of double yielding. Fig.\ref{fig7}(a) presents representative stress-strain curves at the same fixed shear rate as in Fig.6 for an AG with $\beta\epsilon=1.5$ and $\phi=0.6$ for widely variable attraction ranges. Double yielding emerges from a competition between two factors: the bonding length scale defined by the attraction range governing the first yield point, and a larger (deformed) caging length scale which dictates the second yield point. By disentangling the role of these length scales on the nonlinear mechanical response, a deeper understanding of the double yielding mechanism can be elucidated. \\

For widely variable attraction ranges of $a\in\left[0.001,\ 0.5\right]$, where the smallest studied is at least one decade smaller than the cage scale, the stress-strain response after the first yield point closely resembles that of hard spheres. This similarity is particularly evident for the extremely short range case of \textit{a=0.0001} as shown in \textit{SI} Fig. S5. This range corresponds to a bonding length scale of only 1 Angstrom for a micron-sized colloid. This, in principle, can realized by employing colloids coated with special groups that interact via a special (e.g., hydrogen-bonding) chemical mechanism of interparticle attraction. As the attraction range increases, the bonding length approaches the cage scale, leading to a more complex rheological response. In this regime, the cage breakup is delayed to much larger strains, as the bond breaking yield point becomes comparable to the repulsive force caging yield point (see Fig.\ref{fig7}(b)). Further increases in the attraction range diminish the effect of bonding, resulting in a stress-strain curve that converges with that of hard spheres. For $a=0.05$, the stress-strain curve qualitatively mirrors the hard sphere case. Plots complimentary to Fig.7 of the structural and elastic modulus evolution are shown in \textit{SI} Fig.S5.   \\

The overall stress scale increases with decreasing attraction range, particularly the first yield overshoot stress, due to stronger attractive \textit{forces}. The first yield strain associated with bond breakage grows almost linearly with the attraction range (Fig.\ref{fig7}(b)). Therefore, both the first and second yield strains are influenced by the attraction range. These predicted trends are supported by a simulation study \cite{Moghimi2020}, and the predicted behavior of the second yield point is consistent with experiment \cite{Koumakis2011}. {Note that the first yield point, though linearly related to the attraction range consistent with physical intuition, does not quantitatively equal the attraction range which has been roughly seen in some experiments \cite{Moghimi2020,Koumakis2011}. But this is not surprising, since the meaning of spatial range in an exponential potential is different than the polymer-mediated depletion potential in the experimental colloid-polymer suspension \cite{Moghimi2020,Koumakis2011} which is not of an exponential form.  }\\

\subsection{Role of Packing Fraction: From Attractive Glasses to Dense Gels}

We now explore the horizontal trajectory in Fig.\ref{fig3}(a) to study the effect of changing packing fraction on double yielding. We note it has been experimentally observed that moderately concentrated  gels also exhibit the double yielding phenomenon \cite{Koumakis2011,Moghimi2017}. However, the physical mechanism is believed to be different and associated with nonequilibrium cluster formation. It has been proposed that upon straining a gel, intra-cluster and inter-cluster bonds break at different levels of deformation. {This picture was tested in Ref.\cite{Moghimi2017}, where the shear rheology was analyzed for both heterogeneous (with clusters) and homogeneous (without clusters) microstructural states of the colloidal suspension. The heterogeneous system exhibited double yielding, whereas the homogeneous system yielded in a single step.}  In our theory, we analyze only {the latter case of initially equilibrated quiescent} structurally homogeneous systems (no nonequilibrium clusters) {as seems likely most relevant} at the very high packing fractions of present interest. Hence, if the colloid concentration is reduced enough such that cages disappear ($\phi\approx0.44$ for pure hard spheres), we expect a transition from double to single yielding per a dense gel (DG).\\

\begin{figure}[t]
	\includegraphics[width=0.45\textwidth]{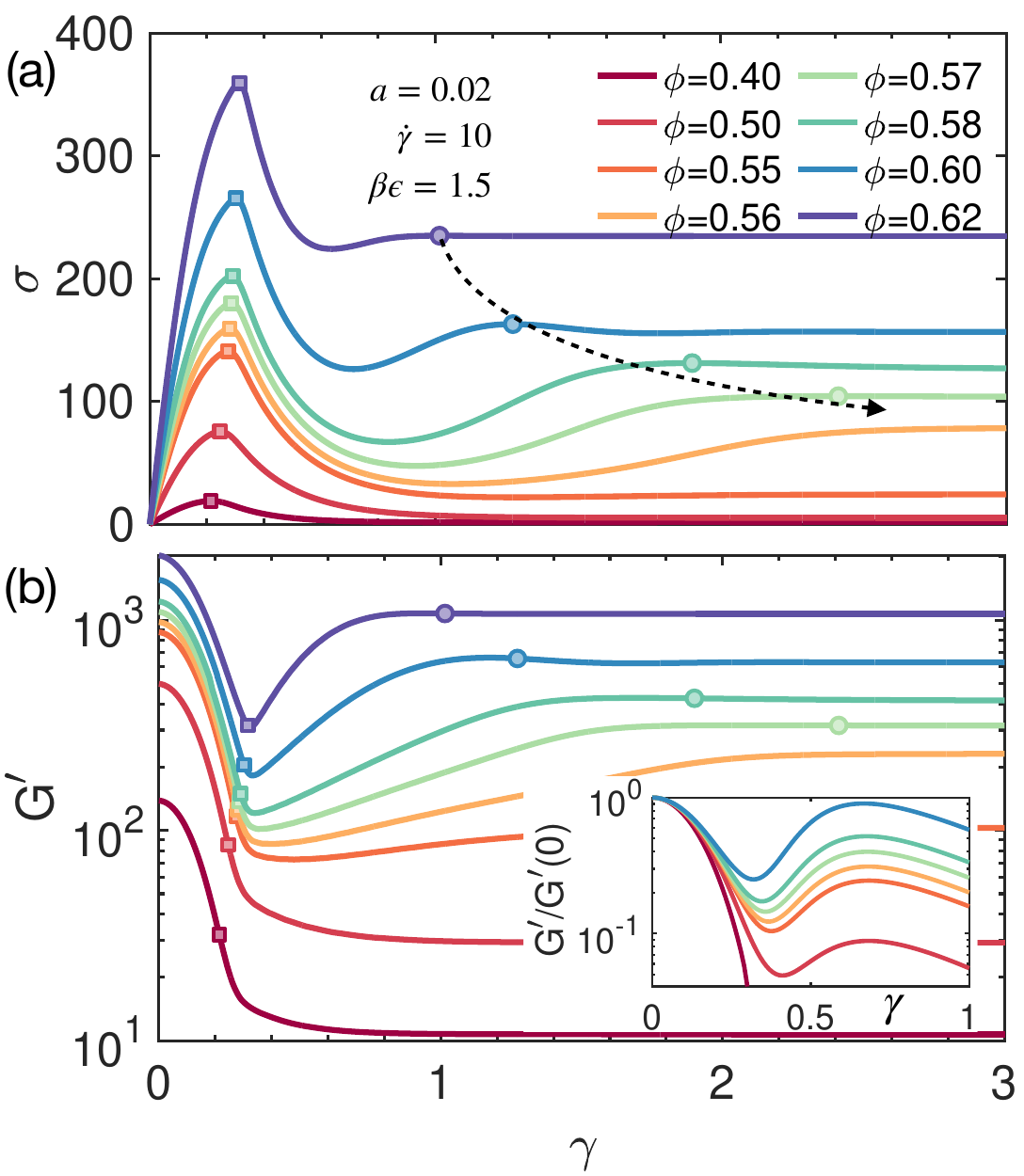}
	\caption{{Step rate} shear rheology of attractive glasses and dense gels: Role of packing fraction. (a) Stress-strain curves for $\phi\in\left[0.4,0.6\right]$. For $\phi\le0.55$, the double yielding behavior disappears, and the cage breaking overshoot shifts outward as packing fraction decreases, while the bond breaking overshoot strain shows minimal dependence on $\phi$. (b) Strain evolution of the elastic modulus for different packing fractions. After the first yield point, the second yield strain shifts outwards (inset). In turn, the absolute magnitude and amount of rebound of the elastic modulus is small, which delays yielding. {Packing fraction increases from bottom to top.}}
	\label{fig8}
\end{figure}
Fig.\ref{fig8}(a) shows representative predictions for the evolution of the stress-strain response with decreasing packing fraction that spans the AG to DG regimes. As packing fraction decreases, the overall stress scale decreases, and the second yield peak overshoot eventually vanishes. The first yield strain remains relatively constant or even slightly decreases as the packing fraction grows, underscoring the dominance of physical bonding for this feature. In strong contrast, the second yield strain increases with decreasing packing fraction, reflecting the enhanced ductility of the system once the physical bonds are broken. Moreover, the caging length scale modestly increases, thereby providing more separation of the bond and cage breaking processes.\\

Fig.\ref{fig8}(b) shows the corresponding evolution of the elastic shear modulus. As packing fraction is decreased, the rate of deformation-induced constraint buildup, and thus the amount of further deformation, slows down after bond breakage, resulting in the second yield point shifting to higher strains. Further insight follows from the elastic modulus evolution predicted in the limit of pure structural deformation via strain (no direct involvement of stress) as shown in Fig.\ref{fig5}. The results in the inset of Fig.\ref{fig8}(b) shows that larger deformation is needed to reach the same level of modulus rebuild at lower packing fractions. This reduces the elastic modulus and the degree of non-monotonic upturn or “rebound” towards the steady state, thereby providing a physical mechanism for the observed slower deformation after the first yield point. For the lowest packing fraction $\phi=0.4$ system, such a rebound behavior is not predicted due to the absence of HS repulsive force induced long-lived cages. Our prediction of increased ductility with decreasing packing fraction is supported by experiment \cite{Koumakis2011}. The corresponding plots of the theoretical structural and alpha time evolutions are shown in \textit{SI} Fig.S6. \\

\subsection{Role of Variable Strain Rate}
\begin{figure}
	\includegraphics[width=0.45\textwidth]{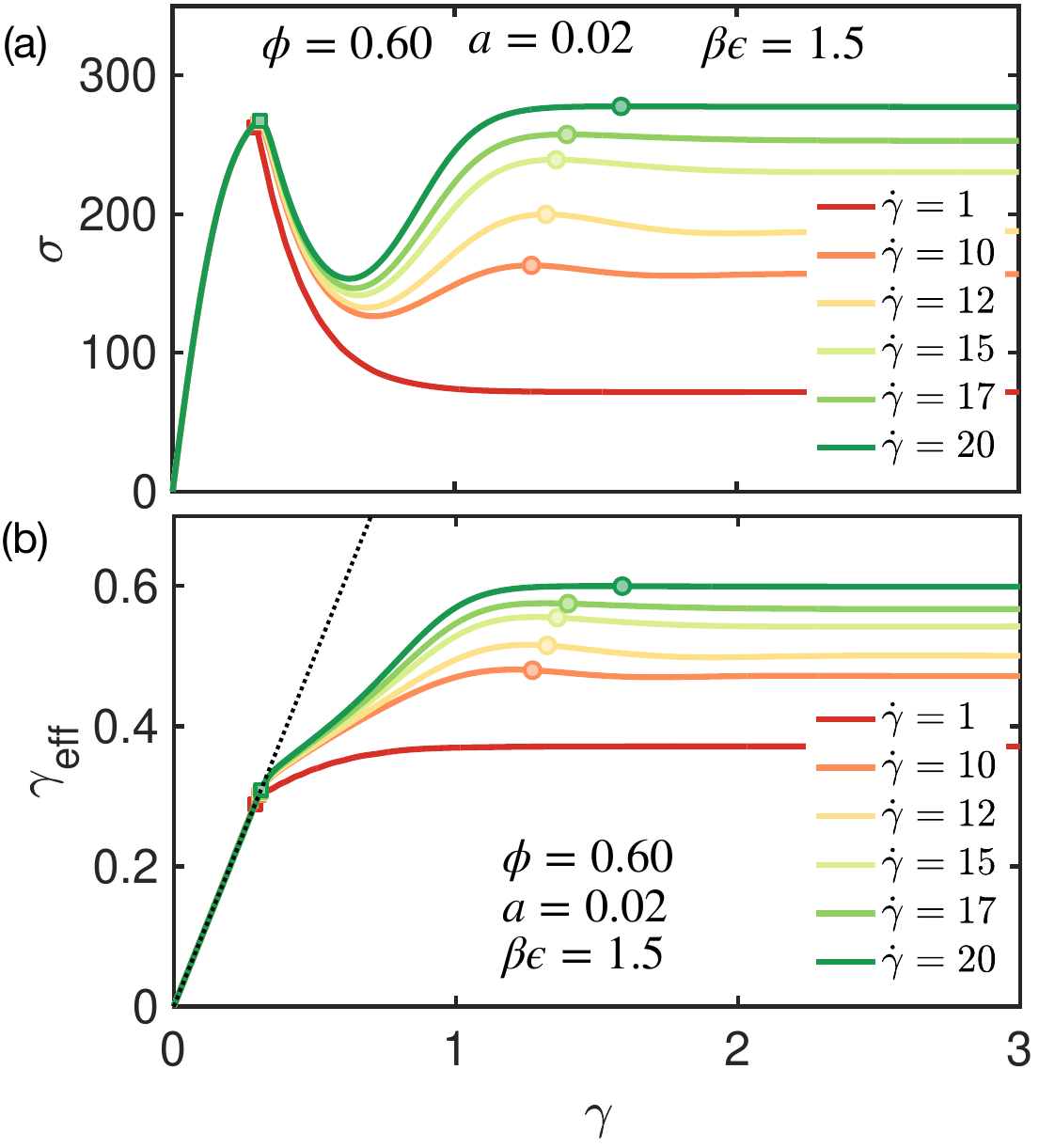}
	\caption{ {Step rate} shear of attractive glasses: Role of strain rate in attractive glasses. (a) Stress-strain curves for a fixed attraction strength $\beta\epsilon=1.5$ and range $a=0.02$ at $\phi=0.6$ for six dimensionless strain rates. The first yield point is weakly affected by strain rate, while the second yield stress decreases and eventually vanishes as shear rate becomes smaller. (b) The corresponding structural evolution parameter. {Strain rate increases from bottom to top.} }
	\label{fig9}
\end{figure}

Changing strain rate directly probes the role of the deformation-dependent \textit{activated} structural relaxation on the rheology. Fig.\ref{fig9}(a) shows representative stress-strain curves for an AG that exhibits double yielding with $\beta\epsilon=1.5$ (far from the glass melting regime), $a=0.02$, and $\phi=0.60$ at strain rates that vary by the large factor of 20. The latter results in major changes in mechanical response, including a striking transition from ductile double yielding to very brittle single yielding with decreasing strain rate. The bonding overshoot (first peak) depends little on strain rate, but does shift in the expected direction of increasing strain and stress as shear rate grows. On the other hand, the second yield overshoot is strongly sensitive to deformation rate. Its behavior spans a wide range, from non-existence a small strain rates, to a huge second overshoot, to eventually becoming more intense than the bond breaking overshoot. The second yield strain also grows with strain rate. \\

The above rich behavior arises from the system having different amounts of time to dynamically relax on the experimental timescale set by the inverse shear rate. Larger strain rates will allow less structural and stress relaxation to occur. Thus, the cages can become more elastically deformed and store more stress. This extra deformation also leads to a slight increase in the second yield strain with increasing strain rate. On the other hand, more relaxation for smaller strain rates suppresses elastic structural deformation, and the net deformation never reaches the level required for deformation-induced constraint built up. This effect can be seen in Fig.\ref{fig9}(b). Thus, the physical bonds are never completely broken. Our predictions of a different behavior of a bonding only yield overshoot for small strain rates, and a dominant second cage breakup overshoot for large strain rates and the significant effect of strain rate on the location of the second yield strain, have been experimentally observed in dense attractive colloidal suspensions \cite{Koumakis2011}.{We note that in ref.\cite{Moghimi2020} the first yield strain was also seen to modestly increase in magnitude with strain rate in experiments, but not in the complementary Brownian simulations. The authors attributed this to the lack of hydrodynamics interactions in the simulations. In Fig.\ref{fig9}(a) we also do not find any significant increase of first yield strain with shear rate. This is consistent with the Brownian simulations \cite{Moghimi2020}, and logically consistent with our model not taking into account hydrodynamic interactions beyond the Stokes-Einstein single particle diffusivity that sets the elementary timescale. Of course, there are other complications in experiments not in our theory model nor in the smooth sphere model simulated. For example, micro-roughness associated with thin brush layers grafted to the colloid surface, and the fact that in experiments the inter-colloid attraction arises from polymer-mediated depletion interaction, in contrast to how attraction is modeled in simulation \cite{Moghimi2020} and our theory.  }\\

Results for the corresponding structural evolution parameter $\gamma_{eff}$ as a function of accumulated strain $\gamma$ are shown in Fig.\ref{fig9}(b). They reinforce the key deductions above concerning the effect of strain rate on the stress-strain curve. The initial response is associated with bond softening and breaking which depends weakly on strain rate. The reason is the enormously large (unmeasureable)  quiescent relaxation time limits the effect of deformation-induced barrier reduction. After the first bond breakage yield point, the structural alpha relaxation time becomes much shorter, and the mechanical response depends on the net elapsed time rather than the net strain. Thus, a larger strain rate results in a smaller relaxation driven effect, a larger slope of the $\gamma_{eff}-\gamma$ curve, and eventually, a larger net structural deformation. The latter drives a larger second overshoot stress, as seen in Fig. \ref{fig9}(b). Complementary plots of the alpha time and elastic modulus that further buttress the discussion above are shown in \textit{SI} Fig.S7. \\

\section{Connections with Experiment and Testable Predictions} \label{Sec5}
{
Throughout the discussion of our new results in section IV we have strived to comment on the qualitative connections to experiments on dense sticky hard sphere Brownian suspensions. Our primary message is we believe the theory based on qualitatively new physical ideas provides the first comprehensive microscopic and mechanistic understanding of the origin of double yielding, how it can be induced as a function of attraction strength, attraction range, shear rate, and packing fraction, and how it can disappear and be replaced by single yielding or two forms of plastic-like response with no stress overshoots. It appears that essentially every major trend observed in experiments is in qualitative accord with our predictions.\\

Beyond this, one can ask about detailed quantitative comparisons with experiment. This is very difficult, given the multiple differences, ambiguities, and complications, both from the theory/model perspective and from the experimental point of view. For example, we employ an effective 1-component model (implicit solvent) with monodisperse sticky colloids interacting via short range exponential attractions (enthalpic bonds) of smooth spheres,  in contrast to experiments that add dilute small nonadsorbing polymers to induce an entropic attraction, the colloids have short brushes on their surfaces, and there is polydispersity and explicit solvent. Thus, how to a priori and uniquely quantitatively map the parameters of our model to reality seems impossible without the introduction of some level of “fine tuning” of parameters which is not our interest here. Having said this, we discuss further below the theory versus experiment question, and do present a quantitative comparison of theory and experiment that we believe strongly supports the essence of our proposed new physics underling the double yielding phenomenon.
}
\subsection{Model Brownian Sticky Colloidal Suspensions}\label{Sec5A}
\begin{figure}[t]
	\includegraphics[width=0.45\textwidth]{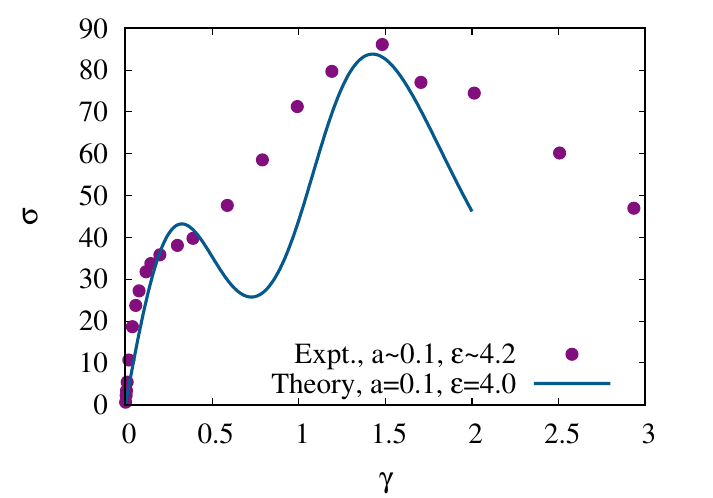}
	\caption{ {Comparison of our theory predictions with step-strain instantaneous stress measurements (points) \cite{Pham2008} on dense attractive PMMA colloidal suspensions (see the figure legend and text for system and parameter details). The dimensionless stress scale in the theory plot is vertically scaled to match the experiment in order to test the theory predictions for the strain dependence. }}
	\label{fig10}
\end{figure}

A modest amount of systematic experimental data on the transient rheological response of well characterized dense model attractive colloidal spheres exists {almost entirely based on the polymer-mediated depletion attraction as the origin of colloidal stickiness. The first observation of attractive glass double yielding was reported by Pham et al.\cite{Pham2008}, who conducted step-strain and large amplitude oscillatory shear (LAOS) experiments \cite{JoshiRheoActa2018,Wagner_Mewis_2021} for very high packing fraction suspensions ($\phi\approx0.60$) and a range of depletion attraction strengths, including the reference HS colloid system. They considered PMMA Brownian colloids (radius $R= 130nm$) mixed with dilute solutions of varying concentrations ($c_p$) of a non-adsorbing polymer of a radius of gyration $R_g= 11nm$. Polymers induce a depletion attraction between colloids of dimensionless short range $a\sim0.09$. The corresponding attraction strength can be estimated \cite{Asakura1954,AOPotentialJCP2022,Chen2004} as $\sim\frac{3}{2}\frac{c_p}{c_p^\ast}\frac{R}{R_g}$ \cite{Chen2004}, where $c_p^\ast$ is the dilute-to-semidilute polymer overlap concentration. For the experimental system, $c_p/c_p^\ast=0.24$, yielding an attraction strength $\beta\epsilon\sim4.2$. \\

Figure \ref{fig10} shows the experimental step-strain instantaneous stress data for the above system that exhibits a double yielding behavior. Also shown are our a priori calculations based on Eq.\ref{eq:StepStrain} with microscopic parameters chosen to connect as well as we can to the specific experimental system. The theoretical result captures all important features of the observations very well. To the best of our knowledge, this is the first microscopic explanation of this striking behavior, and we believe it provides strong support for the key underlying physical ideas of our approach. \\}

In the LAOS measurements, a second plateau in the elastic modulus with increasing strain amplitude was observed, a signature of double yielding. Elastic reentrancy under quiescent conditions was also observed, in agreement with our recent work \cite{Mutneja2024} and Fig.\ref{fig3}. Capturing this subtle quiescent elastic phenomenon requires explicit treatment of attractive forces via the hybrid-PDT force vertex approach, and physically seems crucial to the mechanism of double yielding under highly driven situations. A particularly fascinating observation was the discovery of a second yield point at exceptionally high strain levels of order $100\%$ and higher. The present theory provides an explanation of this observation (per Fig.\ref{fig6}) as a consequence of deformation-induced constraint build-up and the explicit competition between attractive bonds and strongly perturbed repulsive caging.\\

Koumakis and Petekidis \cite{Koumakis2011} expanded on the original studies of PMMA colloids with depletion attractions. They performed {step rate} shear experiments over a wide range of packing fractions $\phi\in\left[0.40-0.60\right]$. The second yield point shifts to larger strains with decreasing packing fraction, as predicted in Fig.\ref{fig8}. They also explored the effect of strain rate on the $\phi=0.60$ system. They found the second yield stress magnitude increased substantially with increasing shear rate and eventually became the dominant overshoot feature, and the second yield point shifts to larger strain with higher strain rates. These striking findings are consistent with the results in Fig.\ref{fig9}. Our suggested physical mechanism is that higher strain rates lead to greater structural deformation and increased deformation-induced constraint built-up, delaying the second yield point and enabling the system to elastically store more stress. The experiments \cite{Koumakis2011} also revealed an increase in the first yield stress with strain rate, a trend not found to date in our model calculation studies.\\

The most recent work by Moghimi and Petekidis \cite{Moghimi2020} on model PMMA colloid depletion systems probed the role of attraction strength. A transition from single to double yielding with increasing attraction strength was observed. Specifically, the bond breaking yield stress increased with attraction strength, and a second yield point emerged for the highest attraction strengths studied at a very large strain of $\sim100\%$, consistent with earlier studies \cite{Pham2008,Koumakis2011}. These trends are consistent with our theoretical predictions in Fig.6. They also performed \cite{Moghimi2020} Brownian simulations that support a linear variation of the first yield strain with attraction range, which is qualitatively consistent with our results in Fig.\ref{fig7}(b).\\

The present theory can broadly probe the controllable large parameter space of packing fraction, attraction strength and range, attraction functional form, and shear rate. This raises future opportunities to design targeted experimental and simulation systems to more deeply test the proposed theoretical ideas and detailed predictions presented in this article and future ones that can be made motivated by new measurements.\\

\subsection{More Complex Systems }
{
The phenomenon of double yielding is often viewed in a general sense as related to  emergent solidity arising from two distinct length scales \cite{Joshi2020}. The simplest system in this context is what we have studied : dense suspensions of sticky smooth hard Brownian colloids which can form physical bonds and cages. Perhaps the next simplest realization is a binary sphere colloidal mixture with a large size disparity \cite{Sentjabrskaja2013} where the different caging length scales are of geometric origin. Double yielding has been observed for sufficiently large size disparities at high concentrations \cite{Sentjabrskaja2013}. Non-Brownian, deformable at the particle level, dense emulsions with short-range attractive forces also exhibit double yielding \cite{Datta2011,Shao2013}, likely driven by the competition of physical bonding and caging but with new aspects associated with particle deformability and soft jamming. At lower packing fractions, colloidal gels \cite{Koumakis2011,Moghimi2017} and pastes \cite{Shukla2015} also display two distinct yielding events generally attributed to elementary bond breakage and yielding on a more mesoscopic cluster level. Addressing these systems in a microscopic approach such as ours is a major challenge that we hope to address in future work, including extending our framework to athermal granular systems.

}

\section{Conceptual Overview}\label{Sec6}
We believe the results presented establish that the theory ideas successfully capture within a single framework the nonlinear evolution of stress and structure during a { step rate} shear. This includes all dependencies of the single or double overshoot effect on attraction strength and range, shear rate and packing fraction, for both hard sphere and sticky dense systems,{ and also the possibility of no overshoot (plastic like yielding) in the re-entrant glass melting region of parameter space. } However, the approach is quite intricate with multiple physical aspects. Thus, in this section we present a qualitative discussion of the origin of the rich behavior obtained in terms of the underlying physics included in the theory, most of which has not been addressed by any prior microscopic statistical mechanical approach. \\

The essential foundation of our present work is prior advances for quiescent thermally \textit{activated} structural and stress relaxation using ECNLE theory plus the hybrid PDT to construct dynamic constraints that explicitly quantifies the different dynamical and elastic consequences of repulsive force caging and attraction-driven physical bonding forces \cite{Mutneja2024}. Both aspects are essential to properly capture the experimentally observed super-Arrhenius growth of the alpha time with packing fraction and attraction strength, and the high packing fraction non-monotonic re-entrant behavior of the elastic modulus and alpha time (and isochronal kinetic arrest boundaries) with attraction strength \cite{Mutneja2024}. These two key variables evolve with the strain-dependent structural deformation and microrheological stress in the nonequilibrium formulation of ECNLE theory in a distinctive manner inherited by the richness of the hybrid-PDT formulation of effective forces in equilibrium. If the latter is ignored and the standard projection of attractive and repulsive forces on density fluctuations is employed per classic applications of ideal MCT, no non-monotonicity of the equilibrium elastic modulus nor a double yielding response in attractive glasses is predicted within our framework.\\

Now, as deformation is applied to an attractive glass, the brittle bonds associated with short range attractions first weaken, mimicking a reduction of attraction strength, while steric cages remain largely intact, requiring higher strains, stress, and structural softening to break or microscopically yield. This sequential destruction of the constraints is what can lead to a double yielding response \textit{via} the predicted non-monotonic evolution of the elastic modulus and activated alpha time. However, a clear double yielding form of the stress-strain curve also requires sufficient dynamical contrast between the effect of deformation on bonding and caging, which depends on the specific fluid density, attractive force magnitude, and external variables such as shear rate.\\

At a technical level, the generalized Maxwell model constitutive equation is a highly nonlinear \textit{self-consistent} equation for the stress quantified by how the elastic modulus \textit{and} alpha time evolve with time-dependent deformation (memory) and how it is functionally coupled to our proposed evolution equation for the nonequilibrium structure. The latter involves an advection driving force that weakens structural correlations in a roughly affine manner, but, crucially, also a relaxation process that tries to drive the system back to equilibrium (crudely ala an aging process) at a rate quantified by the activated alpha relaxation time $\tau_\alpha\left(\sigma,\gamma\right)$. With increasing strain, the stress first increases via the state-dependent elastic modulus, but it also simultaneously relaxes at a strain scale controlled by the effective Peclet number, $\dot{\gamma}\tau_\alpha\left(\sigma,\gamma\right)$, which also opposes the advection-driven deformation. This competition is resolved in a self-consistent manner and introduces a non-affine/plastic component to the structural deformation. \\

At small strains, the alpha time is extremely large, leading to little or no relaxation effects, and thus, the stress increases almost linearly with strain, and the structure deforms almost affinely via advection. In the long time nonequilibrium steady state, structural relaxation must come into balance with elastic deformation. At intermediate strains, there are three possible scenarios of mechanical response based on the timing (or value of accumulated strain) of two aspects that quantify nonlinear elastic and viscous physics: stress relaxation balances stress build-up, and structure relaxation balances structural deformation. (i) If these two aspects happen simultaneously, the stress and structure will monotonically approach a steady state per an ideal plastic response with no stress overshoot. (ii) If stress relaxation matches the stress built up but the structure is still deforming.  This will reduce stress until structural deformation reaches a steady state, and thus an overshoot emerges. (iii) The structural relaxation matches the structural deformation while the stress is still increasing, thereby resulting in no stress overshoot but a structural deformation overshoot. The precise evolution of the elastic modulus and alpha time will determine which of these three scenarios occurs.  \\

 The above discussion largely applies to the nonlinear rheology of both repulsive hard sphere and attractive suspensions. For attractive glasses specifically, the physical mechanism underlying the non-monotonic (“glass melting”) evolution of mechanical and dynamical quantities in equilibrium is inherited under deformation. This introduces new aspects to competing physical phenomena discussed above, resulting in the possibility of a double-yielding response that depends on strain rate, packing fraction, and attraction strength and range. We have verified by numerical calculations that if the direct stress reduction of the elastic modulus and relaxation time is ignored (the microrheological force in the nonequilibrium dynamic free energy in Eq(\ref{eqn:StressAddition}) {is dropped}), but the advected changes of structure with strain and the associated competing relaxation process are retained in a self-consistent manner with stress evolution, then our numerically obtained double yielding predictions are broadly retained. Hence, strain induced structural change and competing relaxation processes that drive the system back to the equilibrium state are key. The presence of the direct effect of stress quantitatively decreases the amplitude of the double yielding overshoots and corresponding value of the yield strain, but it does \textit{not} change the rich qualitative trends. On the other hand, if one ignores \textit{all} effects of coupled stress and strain induced change of structure, then the overshoots vanish (ideal plastic yield response), as previously discussed for hard spheres \cite{Ghosh2023}.\\
 
In all cases, the inclusion of activated structural and stress relaxation, strongly modified by deformation (stress, strain rate, and via structural changes), is crucial in the generalized Maxwell model for our overshoot predictions. Moreover, the coupled self-consistent theory of nonlinear rheology and structural evolution predicts (not effectively assumes, as done previously \cite{Ghosh2023}) the latter process requires higher strains to achieve a steady state, which is critical for predicting a stress overshoot. However, interestingly, per Eq(\ref{eq:TauInftyStrain}) and Fig.\ref{fig5}, if one removes entirely structural relaxation in the constitutive equation ala a non-Brownian system, a 2-step response is still predicted, albeit of a plastic like form (no overshoots)  associated with deformation-induced sequential bond breaking and de-caging. Although there are no stress overshoots in this limit, we believe the so-obtained results capture the most basic effect of 2-step yielding and thus can be viewed as a foundational nonlinear elastic mechanism of deformation-induced sequential breaking of bonds and cages.\\

\section{Summary and Future Outlook}\label{Sec7}

We have formulated, and numerically applied to make predictions and understand diverse experimental observations, a microscopic statistical mechanical theory of the nonlinear rheological response of dense attractive Brownian colloidal systems. The approach is predictive since it is built on casual connections between tunable interactions, structure, dynamics, and mechanics far from equilibrium. It has allowed a unified understanding of the rich evolution under {step rate} deformation of the elastic modulus, stress-strain curve, structural correlations, and activated relaxation time as a function of packing fraction, attraction strength and spatial range, and shear rate. A distinctive change of the mechanical response in repulsive glasses, attractive glasses, and dense gels is predicted. Double yielding is shown to emerge from a competition between strain, shear rate, and attraction strength and range dependent softening of the elastic modulus and relaxation time. Moreover, two other modes of transient response and transformation of a solid to a nonequilibrium flowing liquid are predicted to be associated with a single static yield overshoot where either attractive bonds or repulsive caging dominate, or no overshoot in the “glass melting” re-entrant regime per an ideal plastic response. \\

Concerning the {immediate} future, shear thinning of the viscosity and flow curves can be straightforwardly predicted based on the present work but is beyond the scope of this article. Since the developed physical concepts and theoretical formulation for predicting the coupled nonequilibrium evolution of dynamics, mechanical response and structure are general, { a longer term direction is to extend} our approach to other rheological protocols {for dense Brownian colloidal suspensions.}  This includes nonlinear creep under constant stress, nonequilibrium stress relaxation after an instantaneous step strain, or more generally an interrupted startup {step rate} shear deformation. The latter is exceptionally important in additive manufacturing and processing applications \cite{Truby2016}, where the material is extruded through a nozzle via a stress-induced solid-to-fluid transition, but then evolves to a nonequilibrium solid steady state in the absence of deformation. {Given our approach naturally predicts both elastic moduli and relaxation times under active deformation, extension to treat yielding as probed by the large amplitude and variable frequency oscillatory shear (LAOS) measurements \cite{Pham2008,Koumakis2011,Datta2011} seems also possible, at least at the level of a first harmonic analysis.}\\

{
There are, of course, many other possible effects that could be important that have not been included in our present model and microscopic statistical mechanical approach. One class of effects is pre-shear history, thixotropy, and physical aging which all involve additional complications and challenges for any  microscopic approach. Within the activated dynamics ECNLE theory and generalized Maxwell model frameworks,  physical aging under quiescent conditions and under active deformation, and pre-shear history  effects, have been successfully addressed previously for polymer glasses \cite{Chen2011,ChenPREAging2010}. The ideas developed there may provide a foundation for treating these aspects for complex colloidal soft matter.  \\

Another class of distinct complications includes near and far field hydrodynamic interactions (HI), colloid surface roughness, shear thickening, and dissipative sliding and rolling friction. None of these have been treated within any predictive \textit{microscopic} statistical mechanical approach, and to do so remains an outstanding open challenge. We note that in the cited experiments and Brownian simulations on smooth sticky colloids that motivated our work \cite{Pham2008,Koumakis2011,Moghimi2017,Moghimi2020}, shear thickening and friction effects were not discussed, and Brownian simulations with smooth particles qualitatively captured almost all of the rich experimental behaviors. This fact suggest a minor role for complications such as the above, and the fundamental new physics of double yielding we propose is not tied to HI. We also have focused on rheologically homogeneous systems, and have nothing to say about shear banding. Predictive treatment of the latter at the microscopic statistical mechanical level is an unsolved and daunting problem. In our work, phenomena such as double yielding reflect intrinsic material physics, and is not a consequence of a macroscopic mechanical instability and emergent spatial inhomogeneity \cite{FieldingPRR2022}. We believe this is consistent with shear banding not being invoked in the experimental work we considered \cite{Pham2008,Koumakis2011,Moghimi2017,Moghimi2020}, and its qualitative agreement with Brownian dynamics simulations with no HI and no shear banding \cite{Moghimi2020}. \\

Thus, we emphasize that despite the many potential complications discussed above, what we believe is our successful understanding of essentially all the distinctive trends observed in experimental dense Brownian sticky colloid suspensions suggests these complications are not of zeroth order importance, for either the single or double yielding phenomena.   \\

Another potential future direction concerns dense, but not dense enough to form homogeneous cages, sticky colloid suspensions. When such “gels” are homogeneous, they experimentally exhibit \cite{Moghimi2017} only a single stress overshoot associated with bonding, as we predict. But if the quiescent gel structure involves nonequilibrium clusters, double yielding has been observed experimentally \cite{Moghimi2017,Koumakis2011}. This is not inconsistent with our theory which adopts equilibrated quiescent structural input which does not have mesoscopic “clusters”.  If the structure factor of nonequilibrium sticky colloids with clusters is known from experiment or simulation, it could be used as input to our dynamical and rheology theories, and one could test whether double yielding is captured.  
Finally, all our work is for thermalized Brownian suspensions where deformation-dependent activated processes are important. However, given our prediction in Fig.\ref{fig10} in the absence of relaxation seemingly captures the essence of double yielding in agreement with instantaneous step deformation experiments \cite{Pham2008}, one might wonder if our theoretical framework can be built on to treat non-Brownian or granular materials. This fascinating direction is presently under study. \\
}
%\nocite{*}
\section*{Supplementary Material}
The supplementary material file contains additional figures and their discussion. 
\begin{acknowledgments}
The authors acknowledge support from the Army Research Office via a MURI grant with Contract No. W911NF-21-0146.
\end{acknowledgments}
\bibliography{DoubleYielding}% Produces the bibliography via BibTeX.

\end{document}

\title{Supporting Information for: Microscopic Theory of Nonlinear Rheology and Double Yielding in Dense Attractive Glass Forming Colloidal Suspensions}
\author{Anoop Mutneja}
\affiliation{Departments of Materials Science and Engineering, University of Illinois, Urbana, Illinois, 61801, USA}
\affiliation{Materials Research Laboratory, University of Illinois, Urbana, Illinois, 61801, USA}
\author{Kenneth S.Schweizer}%
 \email{kschweiz@illinois.edu.}
\affiliation{Departments of Materials Science and Engineering, University of Illinois, Urbana, Illinois, 61801, USA}
\affiliation{Materials Research Laboratory, University of Illinois, Urbana, Illinois, 61801, USA}
\affiliation{Departments of Chemistry, University of Illinois, Urbana, Illinois, 61801, USA}
\affiliation{Departments of Chemical and Biomolecular Engineering, University of Illinois, Urbana, Illinois, 61801, USA}

%% Comment out or remove this line before generating final copy for submission; this will also remove the warning re: "Consecutive odd pages found".
%\instructionspage  

\maketitle
\onecolumngrid
\section{Additional Figures and their Discussion}\label{Sec1}
%%% Each figure should be on its own page
\begin{figure}
\centering
\includegraphics[width=1.\textwidth]{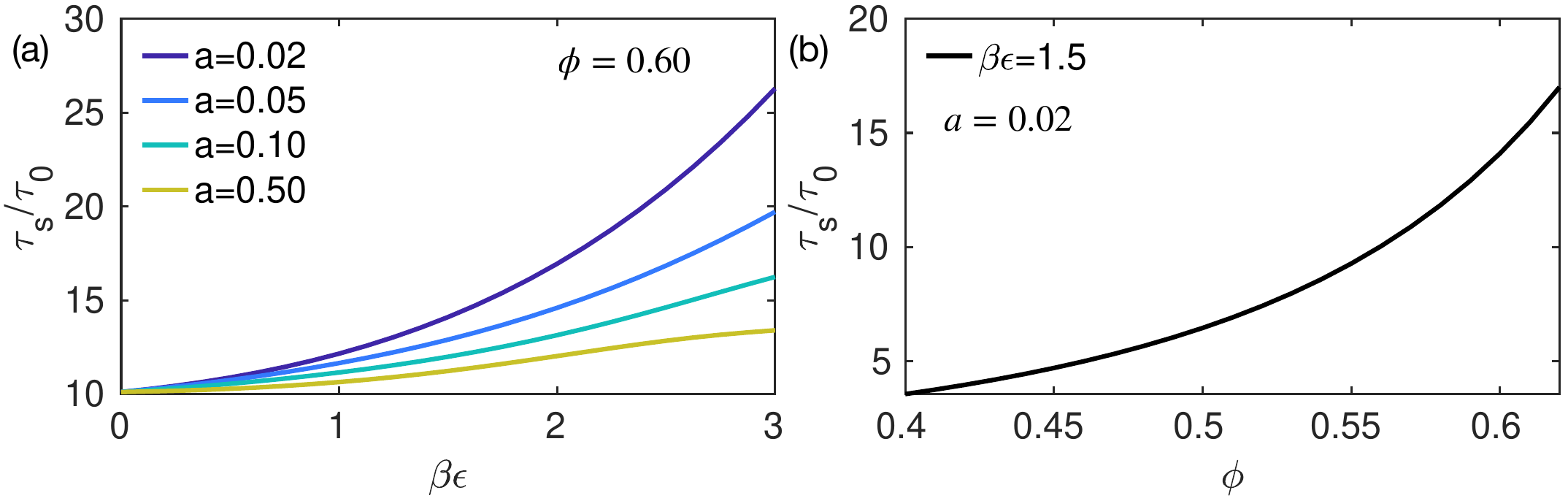}
\caption{Quiescent short time and distance dissipative non-activated process time scale, ${\tau}_s$ over the parameter ranges studied in the main text. (a) ${\tau}_s$ in units of the dilute suspension Stokes-Einstein “bare” time ($\tau_0$) is shown as a function of attraction strength for 4 characteristic attraction ranges at a fixed high packing fraction. Note that ${\tau}_s$ increases modestly by a decade or less, while the activated mean alpha times vary by an enormously larger amount for the shown parameters \cite{Mutneja2024}. (b) Packing fraction variation of ${\tau}_s$ for an attractive glass with $\beta\epsilon=1.5$, and $a=0.02$.}
\end{figure}

\pagebreak
\newpage
\begin{figure}[t]
\centering
\includegraphics[width=1.\textwidth]{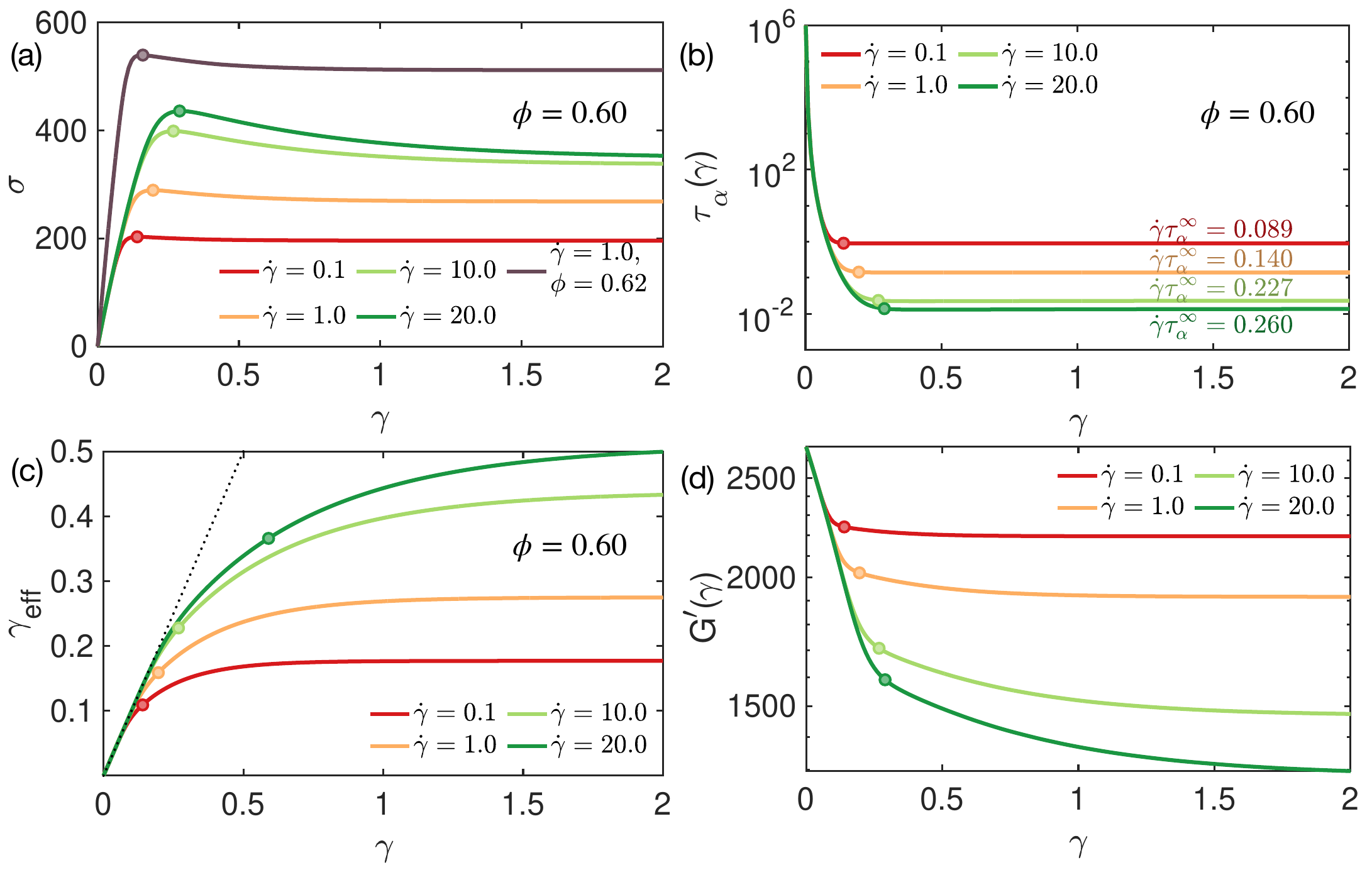}
\caption{Predictions for the step rate shear rheology of hard spheres. (a) Stress-strain curve at ${\phi}={0}.{62}$ (brown) and ${\phi}={0}.{60}$ (other colors). Different strain rate results are shown for ${\phi}={0}.{60}$, with the dimensionless strain rate for ${\phi}={0}.{62}$ fixed at $\dot{{\gamma}}={1}$. The points mark the overshoot location in all the plots. (b) Evolution of the reduced alpha or structural relaxation time (in units of ${\tau}_{s}$) with applied strain at four different strain rates for ${\phi}={0}.{6}$. The alpha time saturates at a steady state value that depends on the dimensionless shear rate, thereby defining a steady-state renormalized Peclet number, ${Pe}\left({\gamma}\right)=\dot{{\gamma}}{\tau}_{\alpha}$,  which lies in the range ${Pe}\in\left({0}.{08}-{0}.{4}\right)$. The alpha time stops evolving significantly with deformation around stress overshoot marking an elastic to viscous transition. (c) The strain evolution of structure in terms of the scalar metric ${\gamma}_{{eff}}$ for the aforementioned systems. At small strains, the bare and effective strains are equal per an affine deformation, ${\gamma}={\gamma}_{{eff}}$ (black dotted line). Deviations then emerge and a steady state value of ${\gamma}_{{eff}}$ is reached that depends on strain rate. (d) Strain evolution of the elastic shear modulus.}
\end{figure}
\pagebreak
\newpage
\begin{figure}
\centering
\includegraphics[width=1.\textwidth]{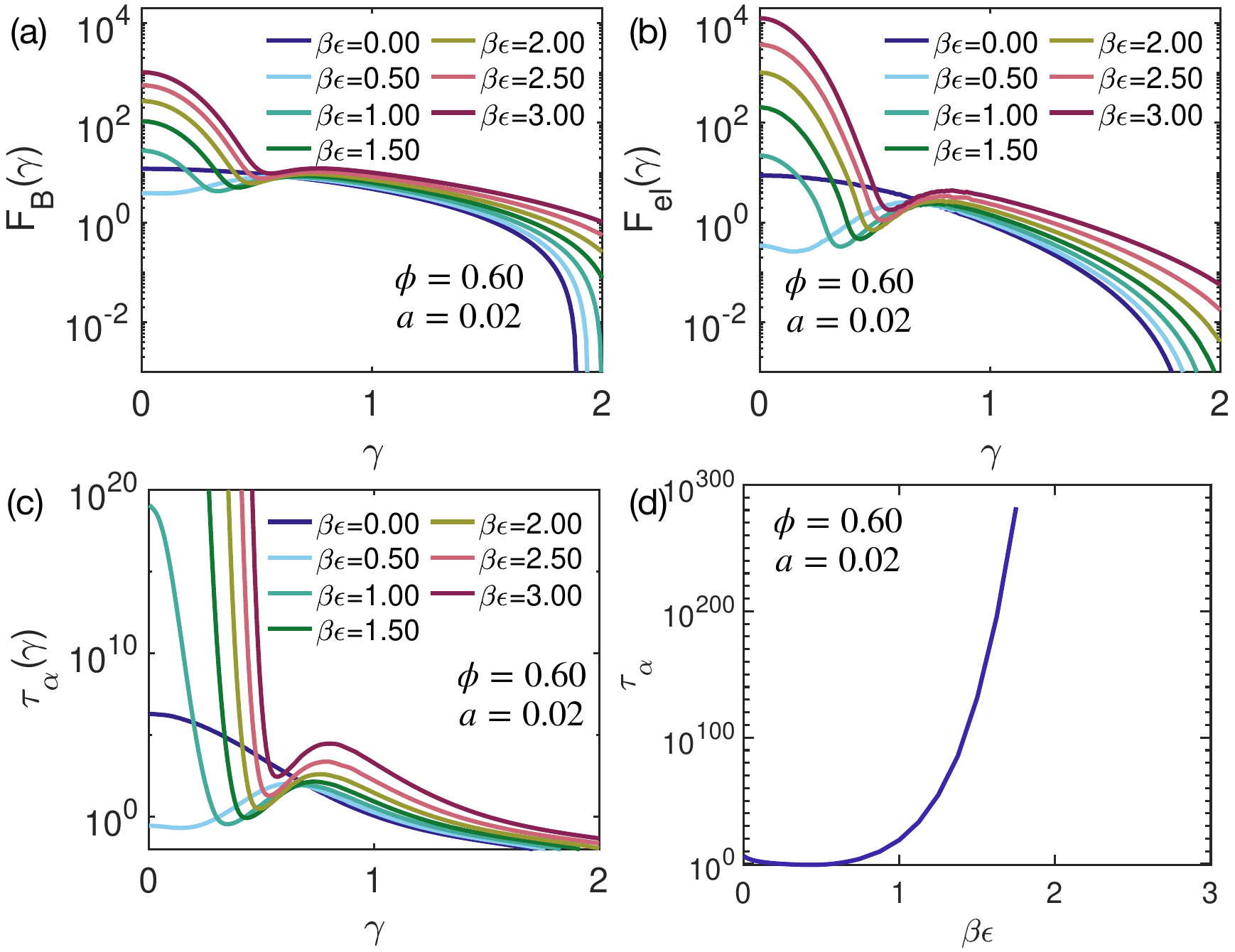}
\caption{\textbf{Effect of strain:} Additional plots that supplement Fig.5 of the main text. Panels (a) and (b) show the local and elastic barrier evolution in thermal energy units, respectively, with externally applied strain for different attraction strengths. Panel (c) shows the corresponding mean alpha time evolution, while the quiescent values shown in panel (d). The off scale part of the curves shown in the figure attain finite values but are astronomically large, and effectively infinite for practical purposes.}
\end{figure}
\pagebreak

\begin{figure}
\centering
\includegraphics[width=1.\textwidth]{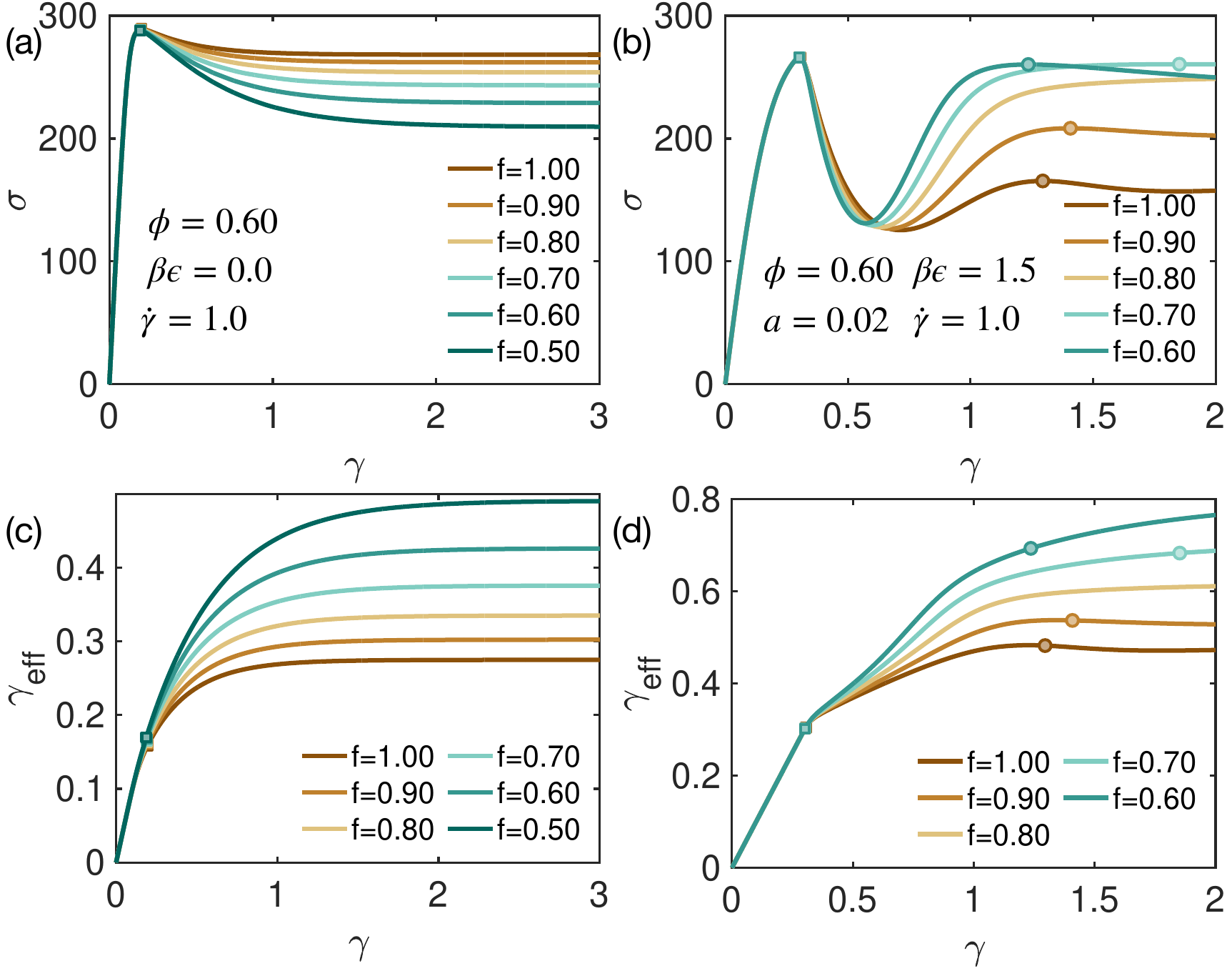}
\caption{\textbf{Effect of mismatch of the stress relaxation and structural relaxation times.} In the main text, the factor ${f}$ was introduced as a quantitative mismatch prefactor between the structural and stress relaxation times; calculations employed ${f}=1$. The structural relaxation time was identified as ${\tau}_{\alpha}/{f}$, where ${\tau}_{\alpha}$ is the Kramers mean first passage time and serves as the stress relaxation time. Here we explore as a representative example the case (which we believe is more physically realistic) when the structural relaxation time is longer than ${\tau}_{\alpha}$, corresponding to ${f}<{1}$. We find our results remain qualitatively the same when varying ${f}$, as shown in this figure. Panels (a) and (c) illustrate the strain evolution of stress and the structural metric parameter for different values of ${f}$ for the hard sphere system. The values at the overshoot remain unchanged, as structural deformation only becomes significant after the overshoot. Reducing ${f}$, or equivalently increasing the structural relaxation time, causes the system to behave more elastically, resulting in greater net deformation (in panel (c)) and thus a lower steady-state stress, along with it requiring larger strain values to reach a steady state. Panels (b) and (d) present similar calculations for an attractive glass with ${\beta\epsilon}={1}.{5}$ and a range $a={0}.{02}$. The first overshoot remains unaffected by a change in ${f}$. However, after the first overshoot, a smaller ${f}$ leads to larger structural deformation. This, in turn, results in a larger deformation induced constraint build up leading to larger second overshoot and steady-state stress, a trend opposite to hard spheres, but a consequence of the same physical reasoning given above. In summary, the effect of relaxation time mismatch on attractive glasses quantitatively alters the ratio of the first to second overshoot stress, while in hard spheres it primarily increases the magnitude of the net overshoot.}
\end{figure}
\pagebreak

\begin{figure}
\centering
\includegraphics[width=1.\textwidth]{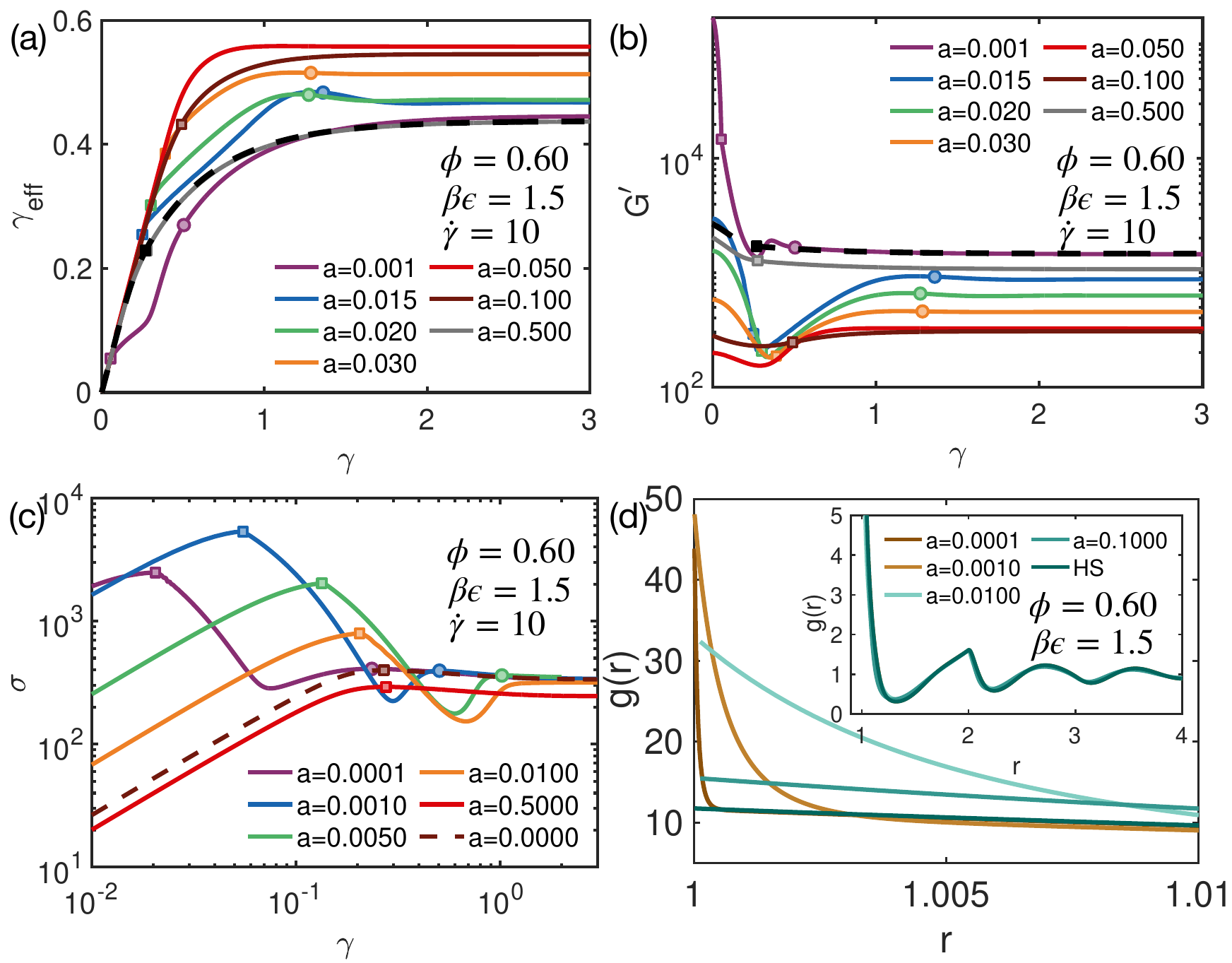}
\caption{\textbf{Effect of attraction range.} These plots supplement the results in Fig.7 of the main text. Panel (a) shows the strain evolution of the structural metric ${\gamma}_{{eff}}$ for systems with different attraction ranges with all other variables fixed. For very short ranges, physical bonds are broken at very small strains, which is followed by a second linear deformation regime that is narrow for short range attractions. The first yield point (marked by a square) increases linearly with attraction range (shown in main text), while the second yield point (marked by a circle) moves outwards from hard sphere value to become extremely large and saturates at $\sim {100}\%$ strain. With further increase in attraction range, the two yield points tend to merge for ${a}\approx{0}.{05}$, and only a single linear structural deformation evolution regime is predicted. The elastic modulus evolution of the same systems is shown in Panel (b), and further supports our arguments. Panel (c) contains the stress-strain plots for an additional system with an extremely short attraction range, ${a}={0}.{0001}$ , plotted in a log-log format. The bonding peak now occurs at a very small strain value, after which the attractive system behaves like hard spheres as shown by the brown dashed line, with the second yield point matching the pure hard sphere value. In panel (d), examples of the quiescent pair correlation function ${g}({r})$ for these short-range attractive systems is shown. For very short-range attractions, ${g}({r})$ closely resembles the pair correlation function of hard spheres beyond a very small interparticle separation of $\sim{a}$. However, for systems with a longer range attraction, the quiescent structure is significantly altered, even at the second caging peak, as illustrated in the inset.
}
\end{figure}
\pagebreak
\begin{figure}
\centering
\includegraphics[width=1.\textwidth]{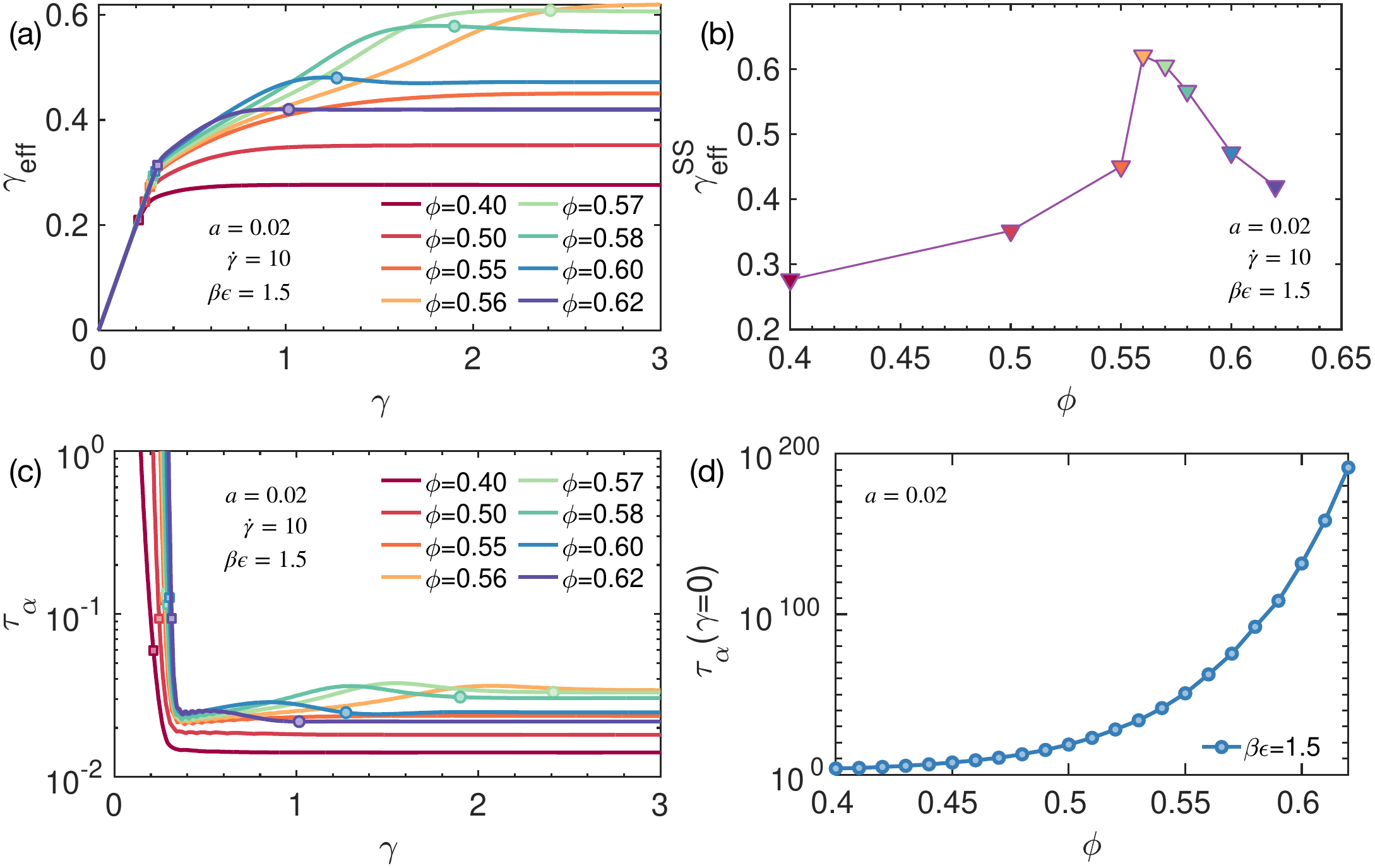}
\caption{\textbf{Packing fraction variations and attractive glasses versus dense gels. }(a) The strain evolution of the structural metric is shown for attractive glass and dense gel systems with varying packing fractions. As discussed in the main text, the second stress overshoot shifts to larger strains with decreasing packing fraction and eventually disappears for ${\phi}\le{0}.{55}$. The double yielding phenomenon arises from the interplay between deformation-induced constraint build-up and structural relaxation as the system approaches a steady state. For the dense gel lower packing fraction systems, constraint build-up occurs at very high deformations (as detailed in the main text), and structural relaxation becomes faster and more dominant before reaching these high deformation levels. Consequently, the system reaches a steady state before the second yield point, resulting in no double yielding and lower overall deformation. As packing fraction increases, the magnitude of deformation-induced constraint build-up increases and occurs at lower deformation levels. For ${\phi}={0}.{56}$, stronger constraint build-up begins to manifest, driving the net deformation to higher values, and leading to a second yield point at very large strain, indicative of increased ductility in the rheological response. Further increases in packing fraction reduce the caging length scale, causing deformation-induced constraint build-up to occur at even smaller deformations. This results in a smaller structural deformation in the steady state, and renders the system more brittle. This behavior is summarized in the steady-state deformation values with packing fraction, as shown in panel (b). Panel (c) depicts the strain evolution of relaxation times for these systems, with a quiescent value shown in panel (d).}
\end{figure}
\pagebreak
\clearpage
\begin{figure}
\centering
\includegraphics[width=1.\textwidth]{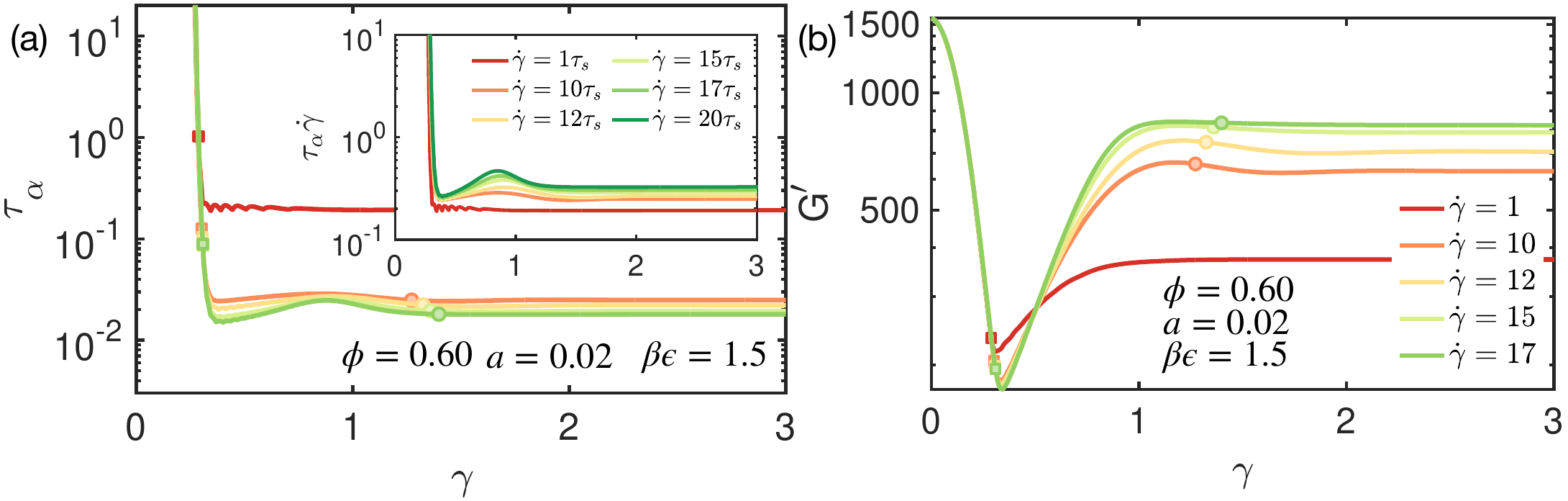}
\caption{\textbf{Effect of strain rate on the continuous start up shear rheology of attractive glasses.} These additional plots complement Fig.9 of the main text and show the strain evolution of the alpha or structural relaxation time in panel (a) and the elastic shear modulus in panel (b) for an attractive glass subjected to different strain rates. The effective Peclet number $\dot{{\gamma}}{\tau}_{\alpha}$ reaches a steady-state value within the experimentally relevant range \cite{Bennin2019} of $\dot{{\gamma}}{\tau}_{\alpha}\in\left[{0}.{1}-{0}.{3}\right]$. The quiescent dimensionless mean alpha relaxation time for the studied system is ${203}.{40}$ }
\end{figure}
%%% Add this line AFTER all your figures and tables

\bibliography{SIDoubleYielding}